\documentclass[superscriptaddress, aps, longbibliography, twocolumn, prb]{revtex4-1}

\usepackage{graphicx}
\usepackage{bm}
\usepackage{hyperref}
\hypersetup{colorlinks=true,citecolor=blue, linkcolor=blue}
\usepackage{physics}
\usepackage{xcolor}
\usepackage{enumitem}
\usepackage{amsmath, amssymb}
\usepackage[normalem]{ulem}
\usepackage{natbib}
\usepackage[inline,final]{showlabels} 

\newcommand{\CZ}{\ensuremath{\textsf{CZ}}}
\newcommand{\Phase}{\ensuremath{\textsf{P}}}

\newcommand{\tmi}{\ensuremath{\mathcal{I}_3}}

\begin{document}

\title{Entanglement phase transitions in measurement-only dynamics}

\author{Matteo Ippoliti}
\affiliation{Department of Physics, Stanford University, Stanford, CA 94305, USA}
\author{Michael J. Gullans}
\affiliation{Department of Physics, Princeton University, Princeton, NJ 08544, USA}
\author{Sarang Gopalakrishnan}
\affiliation{Department of Engineering Science and Physics, CUNY College of Staten Island, Staten Island, NY 10314, USA}
\affiliation{Initiative for Theoretical Sciences, The Graduate Center, CUNY, New York, NY 10016, USA}
\author{David A. Huse}
\affiliation{Department of Physics, Princeton University, Princeton, NJ 08544, USA}
\affiliation{Institute for Advanced Study, Princeton, NJ 08540, USA}
\author{Vedika Khemani}
\affiliation{Department of Physics, Stanford University, Stanford, CA 94305, USA}

\begin{abstract}

Unitary circuits subject to repeated projective measurements can undergo an entanglement phase transition (EPT) as a function of the measurement rate. This transition is generally understood in terms of a competition between the scrambling effects of unitary dynamics and the disentangling effects of measurements.
We find that, surprisingly, EPTs are possible even in the absence of scrambling unitary dynamics, where they are best understood as arising from measurements alone. This motivates us to introduce \emph{measurement-only models}, in which the ``scrambling'' and ``un-scrambling'' effects driving the EPT are fundamentally intertwined and cannot be attributed to physically distinct processes.
This represents a novel form of an EPT, conceptually distinct from that in hybrid unitary-projective circuits.
We explore the entanglement phase diagrams, critical points, and quantum code properties of some of these measurement-only models.
We find that the principle driving the EPTs in these models is \emph{frustration}, or mutual incompatibility, of the measurements. 
Suprisingly, an entangling (volume-law) phase is the generic outcome when measuring sufficiently long but still local ($\gtrsim 3$-body) operators.
We identify a class of exceptions to this behavior (``bipartite ensembles'') which cannot sustain an entangling phase, but display dual area-law phases, possibly with different kinds of quantum order, separated by self-dual critical points. 
Finally, we introduce a measure of information spreading in dynamics with measurements and use it to demonstrate the emergence of a statistical light-cone, despite the non-locality inherent to quantum measurements. 
\end{abstract}

\maketitle


\tableofcontents

\section{Introduction \label{sec:intro} }

The study of out-of-equilibrium quantum dynamics is an exciting research frontier that has seen many important developments in recent years, especially in relation to the dynamics of isolated many-body quantum systems~\cite{mukund, Nandkishore2015, Abanin2019, DAlessio2016, Khemani2019, Sondhi2020, Maldacena2016, Georgescu_2014}. 
Yet more recently, an increased focus on the study of many-body dynamics in \emph{open} systems has emerged, largely motivated by the advent of  `noisy, intermediate-scale quantum' (NISQ) devices~\cite{Preskill2018}. While an ideal quantum computer (or simulator) is a closed unitarily-evolving system, any realistic implementation will have both controlled operations and unintended interactions with its environment, leading to non-unitary, open-system dynamics. 
The study of non-equilibrium -- and possibly non-unitary -- dynamics is a large departure from the usual domain of many-body physics, and requires the development of new tools and paradigms. Ideas from quantum information theory are playing a pivotal role in the development of this new toolkit: in particular (i) \emph{random circuits} have emerged as a versatile tool for the study of many-body dynamics in various contexts~\cite{Nahum2017, Banchi2017, Nahum2018, VonKeyserlingk2018, Rakovszky2018, Gharibyan2018, kvh, Khemani2018, nrh, cdc, Sunderhauf2018, Zhou2019, Hayden2007,Piroli2020, agarwal2019toy} and (ii) universality in the \emph{dynamics of quantum entanglement} has emerged as a novel and incisive paradigm for characterizing many-body systems ranging from electrons in solids to cold atomic gases to black holes~\cite{calabrese2005evolution, zpp, Bardarson2012, hartman2013time, tsunami, kimhuse, islam2015measuring, kaufman2016quantum, mezei2017entanglement, Nahum2017, alba2017entanglement, Rakovszky2019, huang2019dynamics}. 

The study of entanglement dynamics has led to the discovery of novel \emph{entanglement phase transitions} (EPTs), characterized by singular changes in the rate of entanglement growth and/or the entanglement properties of steady states of nonequilibrium many-body systems. A paradigmatic example of an EPT is the many-body localization (MBL) phase transition between a localized phase with logarithmic in time entanglement growth, and a thermalizing phase with algebraic in time entanglement growth~\cite{zpp, Bardarson2012, Serbyn2013}. More recently, a second qualitatively distinct EPT was found in the entanglement dynamics of open quantum systems modeled by circuits of random unitary gates interleaved with local projective measurements~\cite{Li2018,Skinner2019,Li2019}. These works focused on entanglement dynamics along \emph{single quantum trajectories}~\cite{dalibard1992wave, PhysRevA.82.063605}, so that each projective measurement locally collapses the system's wavefunction and a system initialized in a pure state always remains in a pure state (of decreasing norm). 
It was found that individual quantum trajectories exhibit a phase transition as a function of the measurement rate, separating a `disentangling' phase (where the entanglement entropy $S$ obeys an area-law) from an `entangling' phase (where $S$ obeys a volume-law). 

Despite years of sustained effort, the nature and existence of the MBL EPT remains an active and hotly debated area of study~\cite{Pal2010, Luitz2015, KhemaniHuse2017, Imbrie2016, deroeckAvalanche, VHA, PVP, Goremykina_2019, Vidmar2019, AbaninVasseur2019, Panda_2020, Sierant2020, Sels2020}. 
Likewise, the discovery of the unitary-projective EPT was initially a surprising result.
While the existence of the area-law phase was expected (as frequent strong measurements entangle the system to an environment, monogamy~\cite{Coffman2000} prevents different parts of the system from becoming entangled with each other), a robust volume-law phase was not. Its existence conflicts with the intuition that quantum coherence is a delicate resource, unstable to the decohering effects of an environment.
Moreover, while entanglement takes a long time to locally build up and propagate, it can seemingly be destroyed globally by a single measurement.  A very useful perspective on the transition, which clarifies how these issues are sidestepped, is achieved by thinking in terms of quantum information scrambling \cite{Choi2019b,Gullans2019A, Bao2020}: chaotic unitary dynamics tends to hide quantum information in highly nonlocal correlations that are inaccessible to local measurements \emph{i.e.} it forms good quantum error correcting codes (QECCs)~\cite{Hayden2007,Brown2013}.  Local measurements then do not learn much about the state of the system.

In light of this, it is natural to ask just \emph{how} scrambling the unitary evolution must be to protect a volume law phase. A physically relevant case where this question may be probed is for MBL systems, where both information scrambling and entanglement dynamics are logarithmically slow in time~\cite{zpp, Bardarson2012, Serbyn2013}. 
One might expect such a slow scrambling to be unable to compete with measurements performed at a finite rate, leading generically to an area-law phase. 
Surprisingly, we find that this is not true, even in models where the unitary dynamics is strictly non-scrambling (rather than slowly scrambling) -- i.e. the combination of non-scrambling unitary gates and strictly local (single-site) ``unscrambling'' measurements can somehow still furnish a volume law phase! 
We show how these non-scrambling models can equivalently be described as \emph{measurement-only dynamics}, where the unitary gates are discarded altogether, at the expense of introducing multi-site (but still local) measurements. 

With this motivation, the present work adds a \emph{third} novel member to the set of known transitions in entanglement dynamics; by studying \emph{measurement-only} non-unitary circuits, we find EPTs that are qualitatively distinct from both the MBL transition and the hybrid-unitary projective measurement transition. Such measurement-only dynamics are characterized by the ensemble of operators that one is allowed to measure on the system. By varying the measurement ensemble, we are able to obtain and characterize rich dynamical phase diagrams and phase transitions in the steady-state entanglement properties. This paper investigates the properties of this type of dynamics and the nature of the associated QECCs and entanglement transitions. A summary of our results follows.

\subsection{Summary of results}

{\bf 1.} We introduce a class of `measurement only models' (MOMs) where the dynamics entirely consist of projective measurements of a predetermined set of local operators, and find that these models generically support both entangling and disentangling phases. These models enrich the study of many-body dynamics in non-unitary settings in various ways, including:

(a) Showing that entanglement transitions are possible in hybrid unitary-projective circuits even when the unitary dynamics, absent measurements, is not scrambling. This shows that measurements can have an \emph{active} role in the formation of the quantum code that supports the entangling phase.  

(b) Introducing a novel EPT separating volume- and area-law entangled phases in MOMs. This transition is distinct from both the MBL EPT and the hybrid unitary-projective EPT, and adds a conceptually new member to the set of known EPTs. In MOMs, the scrambling and un-scrambling effects are fundamentally intertwined, being produced by the same physical phenomenon (measurement); thus, unlike the unitary-projective case where the balance between scrambling and unscrambling is naturally controlled by the measurement rate $p$, here we find that the operative property is the \emph{`frustration'} of the measurement ensemble. In particular, we find that an entangling phase is the generic outcome when sufficiently long, sufficiently random operators are measured on the system.

(c) Identifying a generic obstruction to the formation of an entangling phase. We introduce an infinite class of models (those with `bipartite frustration graphs') that cannot support a volume-law phase, but rather exhibit distinct area-law phases separated by a critical point pinned by a duality. These models include and vastly generalize a critical point in free-fermion measurement dynamics that has been previously identified~\cite{Skinner2019}. More generally, the area-law phases in these bipartite models can host distinct varieties of quantum order dictated by the symmetries and topology of the operator ensembles. 

{\bf 2.}
We study the properties of the novel QECCs formed in the entangling phase, as well as the time-dependent code properties at criticality. We find that the latter obey a scaling prediction from conformal field theory, lending support to the existence of a statistical-mechanical description of the critical point.

{\bf 3.}
We introduce a new probe of locality and information spreading in dynamics with measurement, which enables us to show ballistic information spreading with a finite velocity in this type of dynamics. This is surprising in light of the non-local, EPR-like behavior of entanglement under projective measurement: the light cone we identity is a statistical, emergent property of the dynamics. This new probe is a non-trivial advance because `standard' measures for information spreading, such as out-of-time ordered commutators, do not readily generalize to the non-unitary setting.

\subsection{Relation to previous work}

The measurement-only models (MOMs) we introduce add a new member to the family of known phase transitions in the dynamics of quantum entanglement, joining the many-body localization (MBL) phase transition and the entanglement transition in hybrid unitary-projective circuits.
The MBL transition has served as a guiding and paradigmatic example for over a decade; there, individual eigenstates of a closed (isolated) system exhibit a sharp change in their entanglement properties as a function of disorder. Hybrid unitary-projective circuits have come to the fore much more recently and broadened the scope of entanglement transitions to the domain of open systems.
It is interesting to note the connection of the latter to quantum error correction thresholds that are studied extensively in quantum information science; for example, the foundational work on fault-tolerant quantum computation with physical locality constraints~\cite{Aharonov00} is a precursor of the contemporary work on entanglement transitions in hybrid circuits. The common focus is on `coherent' dynamics interspersed by `incoherent' processes (noise, or measurement). 
Our work falls outside of this paradigm, by introducing conceptually distinct models that cannot be viewed through the above lens of `coherent'-vs-`incoherent' processes. 

We note that recent works, some simulataneous with ours, have also considered measurement-only dynamics in free-fermion models which host transitions between distinct \emph{area-law} phases~\cite{Skinner2019, Lavasani2020, Hsieh2020, Lang2020}. Our work substantially broadens the scope beyond free fermions, and generically finds both volume- and area-law entangled phases, with a novel transition driven by the `frustration' of measurement ensembles. The existence of the volume law phase is an {\it a priori} surprising outcome in models without scrambling unitary dynamics. 

Finally we remark on connections between this work and other topics in quantum information theory involving measurements.
While it is known that measurements possess the same computational power as quantum gates~\cite{Nielsen2003}, this resource-theoretic equivalence relies on specific protocols, or structured sequences of measurements.
The dynamics we study, on the contrary, feature \emph{spatiotemporally random} sequences of measurements. 
It is crucial to distinguish the possibility of certain outcomes \emph{as a matter of principle}, in specifically tailored circuits, from their realization in a \emph{stochastic} setting -- an important distinction in quantum information science, especially in the theory of fault tolerance~\cite{Aharonov00}.
We also remark on the distinction between measurement-only dynamics and measurement-based quantum computing (MBQC)~\cite{Briegel2009}: in MBQC one is handed an entangled resource state and, by performing measurements in a specific sequence, obtains the (classical) answer to a predefined computational problem, at the expense of destroying the resource state. 
In contrast to MBQC, the measurement-only dynamics we study do not rely on initial resources, are spatiotemporally random, and can produce highly entangled states as their output.

\subsection{Structure of the paper}
The paper is organized as follows.
In Sec~\ref{sec:lbit} we consider the question of entanglement transitions with non-scrambling, MBL-inspired unitary circuits and projective measurements.
This motivates the introduction of measurement-only models (MOMs), which we define in more generality in Sec.~\ref{sec:mom}.
In Sec.~\ref{sec:ensembles} we show numerical results for the entanglement phase diagrams in several MOMs and draw general lesson on their phenomenology, including their critical properties in Sec.~\ref{sec:critical}. 
In Sec.~\ref{sec:bipartite}, we introduce a class of models -- those with ``bipartite frustration graphs" -- which present an obstruction to the existence of a volume-law phase. Instead, such models can host distinct area law phases, possibly characterized by different types of quantum order and separated by novel phase transitions. 
The remainder of the paper focuses on different properties of the volume-law phase and of critical points in the previously introduced MOMs:
Sec.~\ref{sec:code} discusses their properties as quantum error-correcting codes, while Sec.~\ref{sec:loc} focuses on their locality and causality structure. 
We conclude by summarizing our results and pointing to open questions and future directions in Sec.~\ref{sec:discussion}.


\section{Motivation: ``measurement-enabled entanglement'' in $l$-bit circuits \label{sec:lbit}}

An appealing interpretation of the entanglement phase transitions in unitary-projective dynamics is based on the competition between the scrambling effect of unitary dynamics and the ``un-scrambling'' effect of local measurements.
Measurements tend to degrade locally-accessible quantum information into classical bits, while chaotic unitary dynamics tends to scramble, i.e. hide quantum information in nonlocal degrees of freedom, where it cannot be accessed by local projective measurements. 
It is thus tempting to conjecture that the phase of the system is decided by which process happens faster -- the information hiding due to the unitary dynamics or the read-out induced by the measurements. 
Such a scenario yields a critical measurement rate $p_c$ below which the system's steady state has volume-law entanglement. 
This scenario also suggests that by tuning the scrambling rate to zero, one should be able to push $p_c$ down to zero. As we will now see, this is not the case. 

\subsection{Removing scrambling: $l$-bit models}

To test the above scenario, we consider what happens when the unitary dynamics does \emph{not} scramble
at a finite rate.
The many-body localized phase provides an example. Here, entanglement is known to grow only logarithmically in time upon quenching from an unentangled product state~\cite{zpp, Bardarson2012, Serbyn2013}. Despite the slow growth, the entanglement entropy at late times typically saturates to a volume law with a sub-thermal entropy density~\cite{Bardarson2012, Serbyn2013}.
Slow scrambling can be explained using the $l$-bit representation of a fully-MBL Hamiltonian~\cite{Huse2014, Serbyn2014, imbrie},
\begin{equation}
H_\text{MBL} = \sum_i h_i \tau^z_i + \sum_{k \geq 2} \sum_{i_1<\dots<i_k} J_{i_1,\dots i_k} \tau_{i_1}^z \cdots \tau_{i_k}^z
\label{eq:hmbl}
\end{equation}
where the $\tau^z_i$ operators are local integrals of motion ($l$-bits)\footnote{We note that the l-bit Hamiltonian is generally viewed as a rewriting of a more realistic disordered MBL quantum Hamiltonian (with diagonal and off-diagonal terms), so that the l-bit operators $\tau_i^z$ are superpositions of physical spin operators (p-bits) with support in an exponentially decaying spatial envelope around site $i$. However, in what follows, we will simplify our analysis by directly studying the diagonal l-bit Hamiltonian as a model in its own right, with the $\tau_i$ operators treated as strictly local. Such a model still displays log-growth of entanglement, as discussed above.} and the exponentially decaying couplings, $J_{i,\dots j} < e^{-|i-j|/\xi}$, causes a logarithmic growth of entanglement in quenches from generic product states due to slow dephasing between different $l$-bit basis states.

\begin{figure}
\centering
\includegraphics[width=\columnwidth]{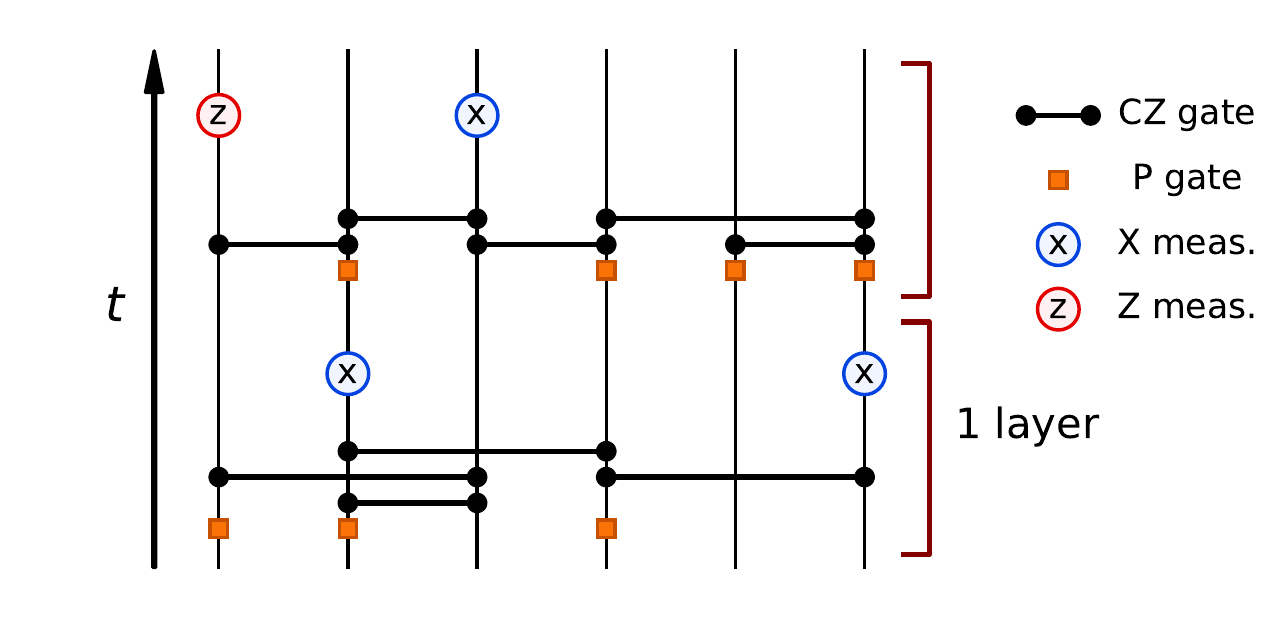}
\caption{Schematic of the $l$-bit circuit with projective measurements.
This circuit has $n=3$: $\CZ$ gates are allowed between qubits $i,j$ with $|i-j|<3$, i.e. nearest and next-nearest neighbors.
All allowed gates happen with probability $1/2$ in each layer.
\label{fig:lbitcircuit}}
\end{figure}

We now ask whether interspersing the above dynamics with local projective measurements yields an entanglement transition.  
%
To address this question further, we focus on an even less entangling model of unitary dynamics: in Eq.~\eqref{eq:hmbl}, we only allow two-body couplings out to a \emph{finite} distance $n$. 
This cut-off gets rid of global scrambling altogether, capping the amount of entanglement to an $O(n)$ size-independent (hence area-law) value upon starting from a non-entangled initial product state. 
Moreover, to facilitate numerical simulations, we consider a toy Clifford circuit $l$-bit model with two gates: 
the two-qubit controlled-$Z$ gate $\CZ_{ij} = e^{-i \frac{\pi}{4} (Z_i - 1) (Z_j-1)}$ ($i<j<i+n$) and the single-qubit phase-gate $\Phase_i = e^{-i\frac{\pi}{4} Z_i}$.
The system is subject to a layer of unitaries, 
\begin{equation}
\textsf{U}_{l\text{-bit}} = \prod_i \Phase_i^{a_i} \prod_{i<j<i+n} \CZ_{ij}^{b_{ij}}\;,
\label{eq:lbitU}
\end{equation}
with $a_i, b_{ij} \in \{0,1\}$ chosen randomly with probability $1/2$ (notice all the gates commute so there is no need to specify the order in which they act).
Then, for each site, we either measure $X$, measure $Z$, or do not perform any measurement, with probabilities $p_x$, $p_z$ and $1-p_x-p_z$ respectively.
The whole process is iterated until a steady-state distribution of entanglement is reached.
This setup is illustrated in Fig.~\ref{fig:lbitcircuit}.

Notice that $Z$ measurements create distentangled $l$-bits which commute with the unitary dynamics; the only way a measured $l$-bit can again become entangled with the rest of the system is by being measured in the $X$ direction first. Hence the dynamics with $p_x = 0$ and $p_z>0$ trivially leads to a product state as soon as every site is measured once. 
Measuring in the $X$ basis, thus breaking the conservation law, is necessary to obtain any nontrivial steady state.

\begin{figure}
\centering
\includegraphics[width=\columnwidth]{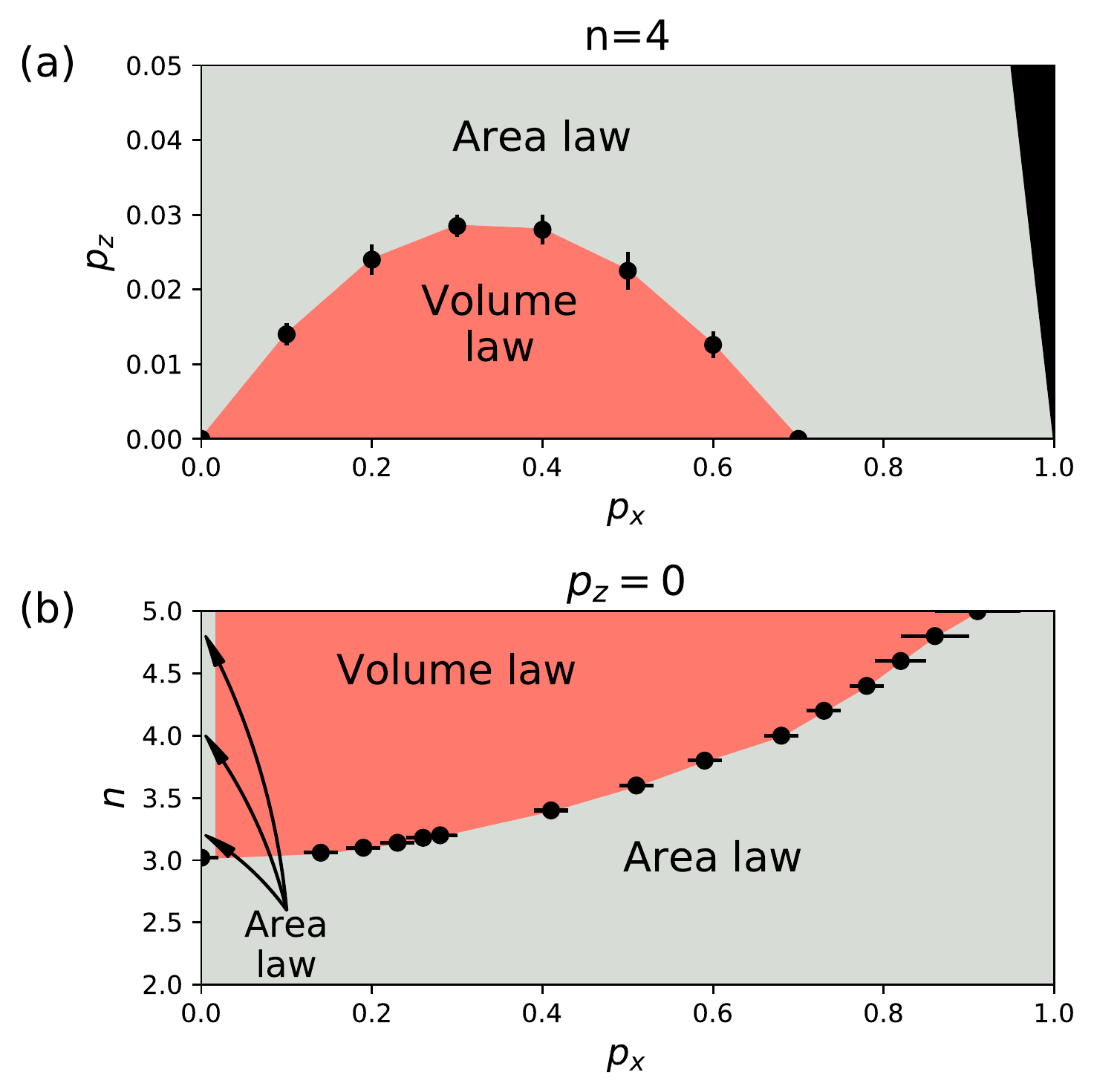}
\caption{Entanglement phase diagram of the $l$-bit unitary-projective dynamics as a function of range $n$ and probabilities $p_x$, $p_z$ of projective measurements in the $X$ and $Z$ basis. 
The black dots represent crossings in the tripartite mutual information (see Sec.~\ref{sec:critical}) obtained from numerical stabilizer simulations of systems of $L\leq 512$ qubits evolved for time $T=4L$, averaged over at least 100 random realizations.
(a) Fixed range $n=4$.
For $0<p_x\lesssim 0.7$ there is a volume-law phase robust to the introduction of sufficiently infrequent $Z$ measurements.
(b) Measurements in the $X$ basis only ($p_z=0$). 
A volume-law phase exists for range $n\gtrsim 3.05$ (see App.~\ref{app:mom_lbit} for the continuation of the models to fractional range $n$). The $p_x=0$ line is always area-law, regardless of $n$. 
\label{fig:lbitphasediag}}
\end{figure}

Surprisingly, we find that the above model (which, to reiterate, has area-law entanglement at both $p_x=0$ \emph{and} $p_x=1$) admits a volume-law phase. Fig.~\ref{fig:lbitphasediag}(a) shows the phase boundary in the $p_x$, $p_z$ plane for the model with range $n=4$.
For $0<p_x\lesssim 0.7$ the model is in a volume-law phase which is robust to the insertion of sufficiently infrequent $Z$ measurements.
A similar picture holds for all $n>3$, with an increasingly robust volume-law phase, see Fig.~\ref{fig:lbitphasediag}(b). 
(The data include fractional values of $n$; see App.~\ref{app:mom_lbit} for a definition of the associated circuit). 
Considering now measurements in the $X$ basis only ($p_z=0$) we find that the model with $n=3$ is area-law for any $p_x$.
A volume-law phase is found for $n\gtrsim 3.05$ -- though precise determination of the critical $n$ requires taking $p_x\to 0^+$ which is subtle. 
We revisit this point from a different perspective in Sec.~\ref{sec:ensembles} and App.~\ref{app:mom_lbit}.

A few comments are in order.
First, this result shows that an interpretation of entanglement transitions in unitary-projective circuits based on the competition between the rates of measurement and unitary scrambling is incomplete: 
entanglement transitions are possible even with a scrambling rate of zero. 
Second, the result does not follow trivially from the fact that the $X$ measurements break the conservation laws in the unitary part of the circuit ($[\textsf{U}_{l\text{-bit}},Z_i]=0$). 
While clearly necessary, this condition is insufficient -- e.g., the models discussed above with $n\leq 3$ remain area-law despite the integrals of motion being broken by measurements.  
Finally, it is remarkable that the interplay of two ingredients that are separately incapable of creating much or any entanglement (the finite-range $l$-bit gates and single-site measurements) can nonetheless yield a volume-law phase. 
This ``measurement-enabled entanglement'' necessitates a new framework. 
In the rest of the article we advance the proposal that such a framework relies on measurements alone.

\subsection{Removing unitary gates: measurement-only models}

\begin{figure}
\centering
\includegraphics[width=\columnwidth]{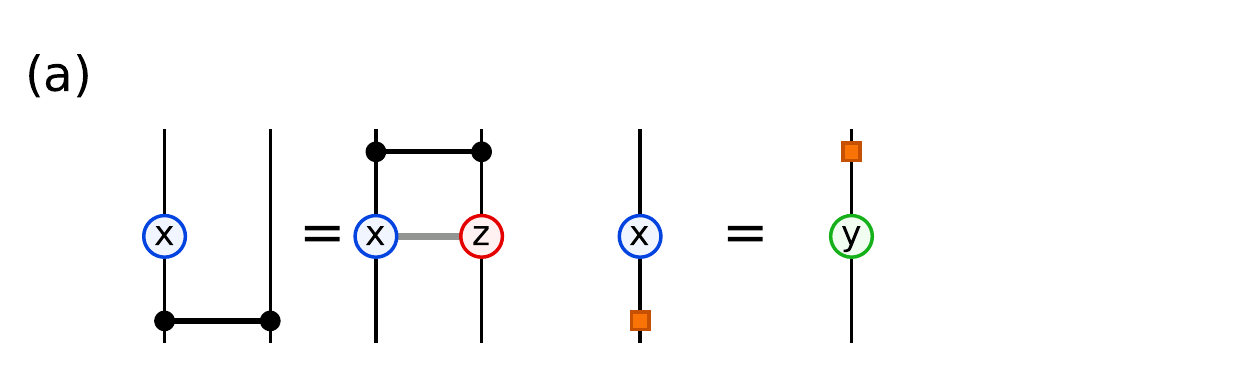}\\
\includegraphics[width=\columnwidth]{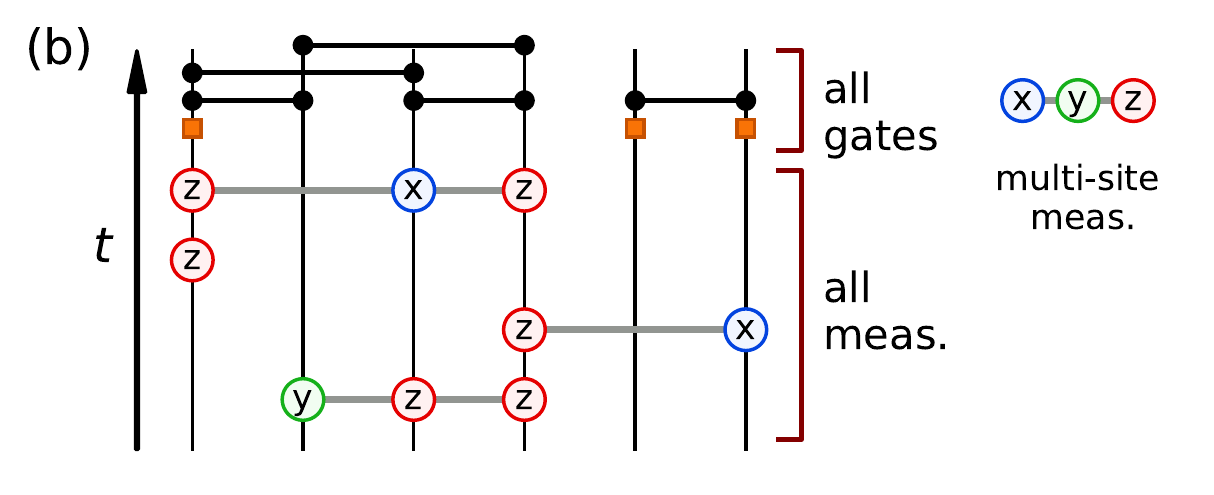}\\
\includegraphics[width=\columnwidth]{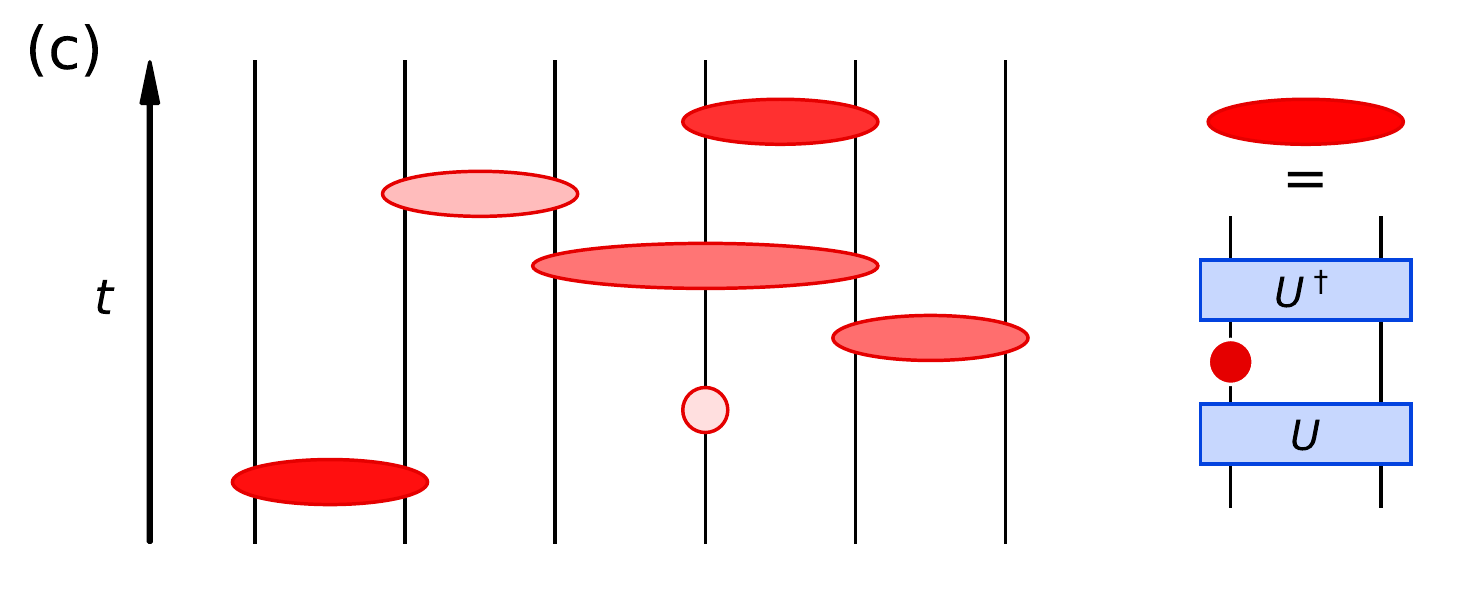}
\caption{(a) Rules for taking $\CZ$ and $\Phase$ gates past single-site $X$ measurements, Eq.~\eqref{eq:trick}.
(b) The $l$-bit circuit from Fig.~\ref{fig:lbitcircuit} after taking all the gates past the measurements. Each single-site $X$ measurement develops ``tails'' of $Z$ operators on either side of maximum length $n-1$ (2 in this case) due to conjugation by $\CZ$ gates.
(c) Sketch of general measurement-only dynamics. Multi-site Pauli measurements (ellipses) are equivalent to single-site measurements conjugated by a unitary gate $U$.
\label{fig:lbitmom}}
\end{figure}

Let us take a Pauli string $O$ and denote the projective measurement of $O$ by $\mu_O$:
$$
\mu_O (\ket{\psi}) = \frac{(\mathbb{I} +s O)\ket{\psi} }{\| (\mathbb I + sO) \ket{\psi} \|},
$$
where $s\in\{+1,-1\}$ is picked randomly according to the usual Born probability, $\text{Prob}(s) = \frac{1}{2}(1+ s\bra{\psi}O\ket{\psi})$.
It is clear from the above definition that the following holds for any unitary $U$ and state $\ket{\psi}$:
\begin{equation}
\mu_O( U \ket{\psi} ) =  U  \mu_{U^\dagger O U}  (\ket{\psi})\;,
\label{eq:trick}
\end{equation}
so that a unitary evolution $U$ followed by a measurement of the operator $O$ is equivalent to first measuring the (typically longer) Heisenberg evolved operator $U^\dagger O U$ followed by the unitary evolution $U$.  
A consequence of this fact, unique to non-scrambling circuits, is that the unitary-projective dynamics can be temporally separated into a unitary part and a projective part that are \emph{both local}.
This is because sliding a layer of $l$-bit gates (Eq.~\eqref{eq:lbitU}) past an $X$ measurement according to Eq.~\eqref{eq:trick} yields an operator with finite support. 
Specifically, since $\CZ_{ij}^\dagger X_i \CZ_{ij} = X_i Z_j$ and $\Phase_i^\dagger X_i \Phase_i = Y_i$ (as sketched in Fig.~\ref{fig:lbitmom}(a)), we have 
\begin{equation}
\textsf{U}_{l\text{-bit}}^\dagger X_i \textsf{U}_{l\text{-bit}} = X_i (-i Z_i)^{a_i} \prod_{\substack{|i-j|<n\\ j\neq i}} (Z_{j})^{b_{ij}} 
\label{eq:lbit_ensemble}
\end{equation}
which is a Pauli string of length at most $2n-1$, characterized by an $X$ or $Y$ operator surrounded by finite ``tails'' of $\mathbb{I}$ or $Z$ operators on boths sides (the exponents $a_i, b_{ij} \in \mathbb{Z}_2$ are as in Eq.~\eqref{eq:lbitU}).
Notice that even when conjugating by several layers of $\textsf{U}_{l\text{-bit}}$, the operator can't grow any longer than this -- the only effect of multiple layers is to change the values of $a_i$ and $b_{ij}$, thus looping through $2^{2n-1}$ Pauli strings of maximum length $2n-1$. 
After taking all the unitary layers to the end of time, $t=T$, we are left with a circuit consisting purely of local multi-site measurements drawn from some finite ensemble of Pauli strings, as in Eq.~\eqref{eq:lbit_ensemble}, followed by a final layer of unitaries (see Fig.~\ref{fig:lbitmom}(b) for an example with $T=2$ layers).
This final layer, despite being the composition of $T$ layers, is in fact equivalent to a single layer having $a_i^{\rm tot} = \sum_{t=1}^T a_{i}(t)$ and $b_{ij}^{\rm tot} = \sum_{t=1}^T b_{ij}(t)$ modulo 2 (which are still uniformly distributed binary numbers).
This can change the entanglement about any given bond by at most $n-1$ bits, which cannot change the entanglement phase (area-law to volume-law or {\it vice versa}). It can thus be safely discarded for our purposes.
This leaves us with a circuit consisting exclusively of local measurements of multi-site Pauli strings.

It follows that the entanglement transition in $l$-bit unitary-projective circuits discussed earlier can actually be understood as the result of projective measurements alone.
While this type of quantum dynamics, known as measurement-induced dynamics, has been considered in the context of quantum information processing and metrology~\cite{Burgarth2014, Pouyandeh2014, Ma2018}, its entanglement properties are largely unexplored. As we have seen above, entanglement phase transitions are possible in this type of dynamics.
From this result it follows that measurements can play both sides in the competition underlying entanglement transitions -- at the same time degrading quantum information to classical bits \emph{and} hiding quantum information from local, accessible degrees of freedom into non-local, inaccessible ones.
This has the potential to either increase \emph{or} decrease the complexity of the state.

We emphasize that while the measurement of a multi-site Pauli string can be viewed as a composition of unitary evolution and measurement of a single-site Pauli operator, these models are \emph{not} the same as the unitary-projective circuit models studied previously~\cite{Li2018, Skinner2019}. 
To wit: the unitary gates before and after the single-site measurement are perfectly correlated, being adjoints of each other (Fig.~\ref{fig:lbitmom}(c));
they would not induce scrambling in the absence of the intervening measurement.

 In the following we introduce a broader class of MOMs, including and generalizing the $l$-bit model discussed above, and study their entanglement properties.


\section{Measurement-only dynamics \label{sec:mom} }

\subsection{Setup}

We define measurement-only dynamics by introducing an `ensemble' $\mathcal E = (P_\alpha, \{ O_\alpha \})$, consisting of a set of Pauli strings $O_\alpha$ and a probability distribution $P_\alpha$ over $\{O_\alpha\}$.
Any such ensemble $\mathcal E$ induces a random dynamics in the following way: at each time step, an $O_\alpha$ is picked according to $P_\alpha$ and measured at a random location in the system; 
doing so updates the state according to the `wavefunction collapse' $\ket{\psi_{t+1}} = \mu_{O_\alpha}(\ket{\psi_t})$;
starting from an initially disentangled product state, this step is iterated until a steady-state distribution of the entanglement over the so-generated ensemble of states is achieved. 

In principle, a 1D system of $L$ qubits has $4^L$ Pauli strings, each of which could be drawn with independent probabilities. 
However, in order to sensibly define phases of matter, we impose the additional requirement of \emph{locality}: we will restrict our ensemble to Pauli strings supported in an interval of length $r$ (the `range' of the ensemble), which does not scale with system size, and specify a probability distribution $P_{\alpha}$ over all $(4^r-1)$ non-identity Pauli strings $\alpha$ in this range. 
While not essential, it is convenient to further assume statistical translation invariance, i.e. that a given Pauli string $O_\alpha$ is measured with equal probability anywhere in the system~\footnote{We emphasize that the assumption of translation invariance applies to the probability distribution underlying the dynamics. Individual realizations of the dynamics will be random in space, as well as time. Measurements arranged in regular spatiotemporal patterns may give rise to different phenomena, as would random measurements with spatially-modulated probability distributions. We leave these directions to future work.}.  In addition, we rescale time as $t \equiv m/L$ where $m$ is the number of measurements that have been performed. 
With this convention, each site is on average subject to $O(1)$ measurements per unit time and the thermodynamic limit is well-defined.

We can view Eq.~\eqref{eq:lbit_ensemble} as one such ensemble:
it has range $r=2n-1$ (maximum length of Pauli strings) and contains $2^{2n-1}$ distinct operators. 
However the underlying unitary-projective dynamics builds in correlations between the probabilities $P_\alpha$ for consecutive measurements, which goes beyond the scope of models defined above. 
Dropping such correlations and taking $P_\alpha$ to be uniform defines a MOM, analyzed in detail in Appendix~\ref{app:mom_lbit}, which is closely related to the $p\to 0^+$ limit of the unitary-projective circuit with measurement rate $p$.

\subsection{Measurements in the stabilizer formalism \label{sec:stab_rules}}

We briefly summarize the update rules for measuring Pauli strings on stabilizer states as they will help in building intuition about measurement-only dynamics. 
A more thorough review is offered in Appendix~\ref{app:stab_rules}.

A stabilizer state is a state of the form
\begin{equation}
\rho = \frac{1}{2^S} \prod_{i=1}^{L-S} \frac{\mathbb I + g_i}{2} \;,
\label{eq:rho_stabilizer}
\end{equation}
where the $\{g_i\}$ are commuting Pauli strings called the stabilizer generators.
If $S=0$ the state is pure, $\rho = \ket{\psi}\bra{\psi}$, with $\ket{\psi}$ the unique simultaneous $+1$ eigenvector of all the $g_i$'s; $S>0$ represents a mixed state.

Measuring a Pauli string $O$ on a state like Eq.~\eqref{eq:rho_stabilizer} can have several qualitatively different outcomes (discussed in detail in Appendix~\ref{app:stab_rules}).
The entropy of a mixed state (Eq.~\eqref{eq:rho_stabilizer} with $S>0$) changes as follows: if $O$ is a `logical operator', i.e. commutes with all $g_i$'s but does not belong to the stabilizer group, it gets added as a new generator and the entropy decreases, $S\mapsto S-1$;
otherwise $O$ is either a stabilizer or an `error' (i.e. anticommutes with at least one$g_i$), and the entropy is unchanged.
In all cases, $O$ itself becomes a stabilizer after the measurement is performed (possibly up to a sign), and the other stabilizer generators may have to be updated to ensure commutation with $O$.

\subsection{Simple limits}

To gain some intuition about measurement-only stabilizer dynamics, we begin by considering two extreme limits: (i) the ensemble of $L$ single-site $Z_i$ operators, and (ii) the ensemble of $(4^L-1)$ global Pauli strings other than the global identity, with operators picked from a uniform distribution in both cases.
The former has range $r=1$ and only one species $O_\alpha = Z$;
the latter has range $r=L$ (which violates the assumption of locality) and $(4^L-1)$ distinct species corresponding to all possible non-identity Pauli strings. 

It is convenient to adopt the dynamical purification perspective~\cite{Gullans2019A} to analyze these two cases. In the purification framework, phases are defined based on the ability of the dynamics to purify an initially mixed state, and are closely associated to the entanglement phases in pure-state dynamics.
We thus start from a maximally mixed state $\rho \propto \mathbb{I}$, measure strings from the ensemble, and decide whether the state purifies in $\log(L)$ time (`pure phase', equivalent to the area-law entanglement phase) or remains mixed out to exponentially long times (`mixed phase', equivalent to the volume-law entanglement phase).

In case (i) the system trivially reaches a pure product state in the $Z_i$ computational basis as soon as every site has been measured once, which takes $O(L\ln L)$ measurements or $O(\ln L)$ time, and thus belongs to the pure, or area-law, phase. Unsurprisingly, single-site measurements can only disentangle.

In case (ii), the first measurement (starting with the identity state) always adds one stabilizer generator $g_1$ to the (initially empty) list, and thus removes one bit of entropy.
The second measurement is equally likely to commute or anticommute with $g_1$: it thus takes two attempts, on average, to add a second generator $g_2$. 
Adding $g_3$ to the generators takes on average 4 attempts, and so on -- the purification time scales exponentially with $L$, and thus the dynamics belongs to the mixed phase. 
This is also not surprising as the strings are completely nonlocal, or all-to-all.

What is not clear from these simple examples is whether it is possible to achieve a volume-law phase by measuring \emph{short} Pauli strings (of finite range $r\gtrsim 1$). 
In this case, the first $O(L/r)$ measurements are likely to commute (simply because the measured strings are unlikely to overlap), adding stabilizer generators and partially purifying the initial state.
However, past this point measurements may begin to frequently anticommute with the existing stabilizer generators and a volume law phase where the system purifies exponentially slowly may occur. 
Whether this happens in practice is not obvious:  
while it is known that arbitrary highly entangled states can be produced via measurements only (because general multi-site measurements are universal for quantum computation~\cite{Nielsen2003}), this relies on very special protocols (e.g. gate teleportation~\cite{Gottesman1999} or entanglement swapping~\cite{Zeilinger1993}).
It is not immediately obvious whether a stable volume-law entangled phase can be generated by measurements placed \emph{randomly} in space and time.
This situation is reminiscent of the entanglement transition in unitary-projective circuits: while it is immediately clear in that case that both volume- and area-law entangled states can be constructed (e.g. in the trivial limits of measurement probabilities $p=0$ and $p=1$), it is not obvious, and was indeed a surprising result, that these extreme limits should extend to \emph{phases} separated by a sharp transition at some critical value $0<p_c<1$.

In the present case of measurement-only dynamics, how to interpolate between the limits considered above is not as clear: 
there is no unique knob to tune (like the measurement probability $p$ in unitary-projective circuits), but rather a huge, multi-dimensional landscape of possible measurement ensembles;
with trivial exceptions like the ones examined above, these ensembles are not straightforwardly sorted from ``more entangling'' to ``more disentangling''.
Understanding this measurement-only dynamics in some generality thus requires a new organizing principle.
In the following, we propose and explore a potential organizing principle: the degree of \emph{``frustration''} of the measurement ensemble.

\subsection{Measurement frustration \label{sec:frustration}}

As we just discussed, fully commuting ensembles of measurements invariably ``localize'' the wave function in a simultaneous eigenstate, which is area-law entangled if the measurements are local.
Some level of non-commutativity among measurements is thus necessary to produce an entangling phase.
Non-commuting observables cannot, by definition, be known at the same time; the wave function thus cannot satisfy all the measurements in the ensemble at once.
We refer to this inability to satisfy non-commuting measurements as \emph{`frustration'}~\cite{Chapman2020}.
It is tempting to conjecture that a suitably defined ``degree of frustration'' (a function of the $O_\alpha$ and $P_\alpha$) could predict the entanglement phase of a given ensemble (without resorting to explicit simulation of finite-size dynamics).

To this end, it is helpful to introduce the \emph{frustration graph} of an ensemble of Pauli measurements~\cite{Planat2007,Chapman2020,Zhao2019, Gokhale2019}.
This is a graph whose vertices represent all the operators in the ensemble $\{O_{\alpha,i} \}$, where $i$ refers to the spatial location of the operator, and where two vertices are connected by an edge if and only if the corresponding operators anticommute.
For local dynamics with operators of a finite range, the frustration graph has a quasi-1D structure, having length $L$ and width equal to the number of operator species, with a periodic unit cell (due to translation invariance), see examples in Fig.~\ref{fig:pofl}(a) and \ref{fig:evenodd}.

The adjacency matrix of this graph defines a four-index object (two species indices $\alpha,\beta$ and two position indices $i,j$), $\Gamma^{\alpha,\beta}_{i,j} = 0$ if $O_{\alpha,i}$ and $O_{\beta,j}$ commute, 1 if they anticommute.
Translation invariance implies that $\Gamma$ only depends on the displacement $\ell\equiv i-j$ between operators. 
In the following we refer to $\Gamma^{\alpha\beta}_\ell$ as the \emph{frustration tensor}.

The frustration graph (or tensor) captures crucial information about the dynamics;
in particular, as we show in Appendix~\ref{app:frustration}, the information therein (plus any algebraic dependence between operators in the ensemble) is sufficient to simulate the dynamics and thus determine the entanglement phase.
Graph-theoretic properties or invariants may place constraints on the existence of an entangling phase in a given ensemble. We return to this in Sec.~\ref{sec:bipartite}, where we discuss a result in this spirit on bipartite graphs.


\section{Phenomenology \label{sec:ensembles}}

In this Section we investigate the generic phenomenology of MOMs with the help of numerical simulations. 
The goal is two-fold: to gain insight into the physical mechanism driving the EPT, and to quantitatively investigate its critical properties and compare them to those of unitary-projective circuits.
To these ends, it is helpful to focus on sufficiently ``generic'' models that exhibit all the phases, while also offering simple handles to tune between them.
We introduce one such class of models in Sec.~\ref{sec:factorizable} and present results on their phase diagrams and EPTs in Sec.~\ref{sec:factorizable_phases} and \ref{sec:critical} respectively.

\begin{figure}
\centering
\includegraphics[width=\columnwidth]{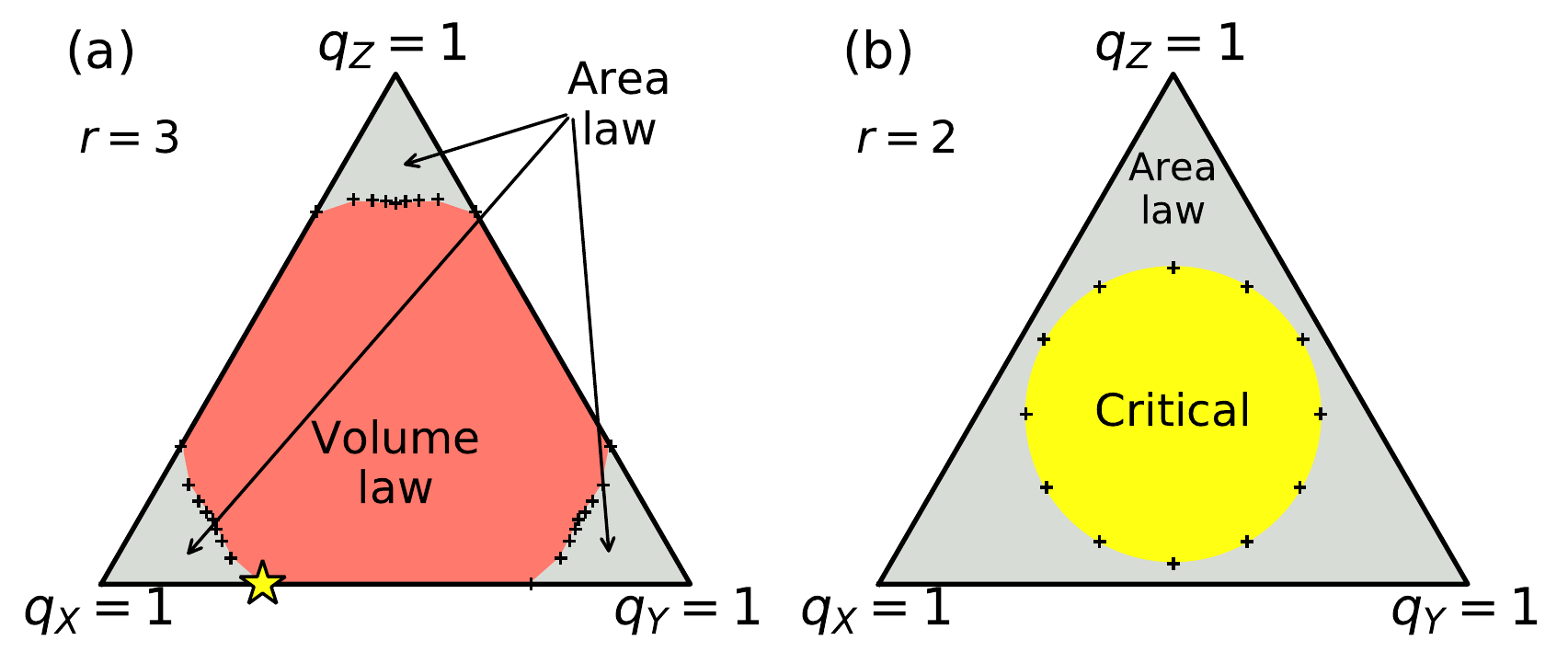}
\caption{Phase diagrams of factorizable ensembles with $q=(0,q_X,q_Y,q_Z)$ (Pauli strings without identities) for ranges $r=3$ (a) and $r=2$ (b). 
The starred point in (a) is the phase transition studied in Fig.~\ref{fig:r3_transition}.
The $+$ symbols are numerical estimates of the phase boundary, obtained from finite-size crossings of the tripartite mutual information $\tmi$ (Eq.~\eqref{eq:tmi}) in (a), and from the finite-size scaling of half-cut entanglement entropy in (b). 
Data was obtained from stabilizer numerical simulations of systems of up to $L=512$ qubits, averaged over between $10^2$ and $10^4$ realizations depending on $L$. Only the wedge $q_X>q_Y>q_Z$ was simulated, with the rest of the phase diagram obtained by symmetry.
\label{fig:triangles}}
\end{figure}

\subsection{Models: ``factorizable'' ensembles  \label{sec:factorizable}}

Measurement-only dynamics as introduced in Sec.~\ref{sec:mom}, even under assumptions of locality and statistical translation invariance, produces a wide parameter space of models -- a hypothetical entanglement phase diagram on range-$r$ MOMs would be $(4^r-1)$-dimensional, which is prohibitive already for $r=2$. 
At the same time most of these dimensions are likely unimportant, and it is crucial to find ways to describe generic measurement ensembles with few parameters. 
We do so by introducing `factorizable' ensembles: sets of Pauli strings (of a fixed length $r$) made from an underlying probability distribution over \emph{single-site} Pauli matrices,
\begin{equation}
O_{\boldsymbol \alpha} = \bigotimes_{n=1}^r \sigma_{\alpha_n}
\implies 
P_{\boldsymbol \alpha} = \prod_{n=1}^r q_{\alpha_n}
\label{eq:factorizable_def}
\end{equation}
where $\boldsymbol{\alpha}$ is a string of Pauli matrix labels $\alpha_n \in \{0,X,Y,Z\}$ and $q$ is a probability distribution over the four single-site Pauli matrices.
This structure reduces the space of models from $O(4^r)$ dimensions to just three dimensions -- the single-site $q_X$, $q_Y$ and $q_Z$ probabilities, which live in a tetrahedron, $0\leq q_\alpha \leq 1$ and $0\leq q_X+q_Y+q_Z\leq 1$. 
Dropping the identity (which tends to increase commutativity, thus likely pushing the dynamics towards area law entanglement) further reduces the phase diagram to the triangle $q_X+q_Y+q_Z = 1$.
The average probability of anticommutation between two measurements is controlled by the vector of probabilities $\mathbf q = (q_X,q_Y,q_Z)$.
More precisely, as shown in Appendix~\ref{app:frustration}, the level of anticommutation is controlled by the distance from the center of the triangular phase diagram, $\delta q \equiv \|\mathbf{q}-\mathbf{q}_0\|$, with $\mathbf{q}_0 = (1,1,1)/3$ the center (all non-identity Pauli matrices equally likely).
Large $\delta q$ means less anticommutation, with the corners of the triangle ($\delta q = \sqrt{2/3}\simeq 0.82$) corresponding to fully commuting measurements.

\subsection{Entanglement phases \label{sec:factorizable_phases}}
We now discuss the phases of the factorizable ensembles for different spatial ranges. 

\emph{Range $r=3$}: We begin by considering ensembles with range $r=3$, which include all $27$ 3-body operators of the form $\sigma_a\otimes\sigma_b\otimes\sigma_c$ with $a,b,c\in\{X,Y,Z\}$, picked with probability $q_a q_b q_c$, and measured at a random location in the system (on 3 consecutive sites).
By simulating this model numerically with the stabilizer method we find that a large part of parameter space belongs to a volume-law entangled phase, see Fig.~\ref{fig:triangles}(a).
Interestingly, the phase boundary is approximately circular: volume-law for $\delta q < \delta q_c$, area-law for $\delta q > \delta q_c$, with $\delta q_c \simeq 0.52$. 
As we noted earlier, $\delta q$ controls the anticommutation probability; therefore this circular phase boundary separates \emph{more frustrated} models (interior, volume-law) from \emph{less frustrated} models (exterior, area-law).
This suggests that measurement frustration is indeed the mechanism supporting the volume-law phase and driving the EPT in these models.

\emph{Range $r>3$}: As the range $r$ is increased, the volume-law phase takes up a progressively larger fraction of the phase diagram (not shown); 
the corners remain trivially area-law for arbitrarily large $r$, but the extent of the area-law parameter space shrinks. 
Focusing on the $q_X+q_Y=1$ ($q_Z=0$) side for simplicity, we find from numerical simulations of $3\leq r \leq 20$ that the critical value $q_{X,c}$ obeys
\begin{equation}
r\simeq \frac{k}{2q_{X,c} (1-q_{X,c})} \equiv \frac{k}{2/3-\delta q^2},
\label{eq:dilute}
\end{equation}
with $k\simeq 1.16$. 
Thus the critical contour approaches the corners of the triangle ($\delta q = \sqrt{2/3}$) with increasing $r$: longer strings generically lead to a volume-law phase, unless they are fine-tuned to be highly commuting.
More specifically, taking $r\to\infty$ and $q_{X}\to 0$ concurrently along the critical line (Eq.~\eqref{eq:dilute}), the probability of sampling the uniform string made entirely of the majority Pauli species (in this case $Y^{\otimes r}$) is $q_Y^r \simeq \left(1-\frac{k}{2r}\right)^r \to e^{-k/2} \simeq 0.56$. 
Thus at criticality a finite fraction of the measurements (in fact a majority) are of the form $Y^{\otimes r}$. 
These operators are mutually commuting and define the stabilizers of a QECC, $\{A_j = Y_j\cdots Y_{j+r}\}$.
This fact points to a qualitative picture for the area-law phase: measurements in the ensemble break up into a code (the $\{A=Y^{\otimes r}\}$ operators in this case) and ``errors'' (all the other operators);
frequent measurements of the QECC stabilizers $\{A_j\}$ constantly remove the ``errors'' before they have a chance to spread and build up any entanglement beyond an area-law.

\emph{Rage $r=2$:} We treat the $r=2$ case separately because it displays qualitatively different phenomenology:
we find no sign of a volume-law phase; instead, we see evidence of a \emph{critical} phase in a circular region around the center of parameter space, as shown in Fig.~\ref{fig:triangles}(b).
The sides of the phase diagram, e.g. $q_Z=0$, map to free fermions~\cite{Cao2019, Nahum2019}:
the Pauli string species $\{X_0 X_1, X_0 Y_1, Y_0 X_1, Y_0 Y_1\}$ are equivalent, under Jordan-Wigner transformation, to $\{i\gamma_{1}\gamma_{2}, 
i\gamma_{1} \gamma_{3}, 
i\gamma_{0}\gamma_{2}, 
i\gamma_{0}\gamma_{3} \}$
where $\gamma_{2j}$, $\gamma_{2j+1}$ are the two Majorana fermion operators on site $j$.
Consistent with the fact that free-fermion dynamics with measurement cannot sustain a volume-law entangled phase~\cite{Fidkowski2020}, we find that the edges of the phase diagram are entirely in the area-law phase. 
For this model, we also see no critical points between area-law phases, unlike other free-fermion measurement-only models~\cite{Nahum2019, Lavasani2020, Hsieh2020, Lang2020}.
The interior of the phase diagram consists of a 9-operator ensemble $\{(X/Y/Z)_0 (X/Y/Z)_1\}$ which does not map to free fermions. 
Numerically we find that the area-law phase identified at the boundary extends in the interior, see Fig.~\ref{fig:triangles}(b);
however, while we can conclusively rule out a volume-law phase anywhere in the interior, the system appears to enter a \emph{critical phase} as $\mathbf{q}$ approaches the center of the triangle $\mathbf{q}_0 = (1,1,1)/3$.
In this phase, we find that the entanglement entropy diverges logarithmically with system size, $S\sim K\ln \ell$.
Though area-to-critical phase boundaries are hard to locate accurately, we find a phase boundary consistent with $\delta q = \|\mathbf{q}-\mathbf{q}_0\| \simeq 0.36$.
Inside this circular contour, the half-cut entropy $S(L/2)$ shows no sign of saturation for sizes up to $L = 512$, and the purification dynamics is consistent with a CFT (we present results on this in the context of quantum code properties in Sec.~\ref{sec:code}).

\subsection{Critical properties \label{sec:critical}}

Having established the existence of entanglement phases in these models, it is interesting to ask whether the entanglement transitions are the same as those found in unitary-projective circuits~\cite{Li2018,Gullans2019B,Zabalo2020}. 
To address this question numerically, we use the \emph{tripartite mutual information}, 
\begin{align}
\tmi (A,B,C) & =
S_A+S_B+S_C + S_{A\cup B \cup C} \nonumber \\
& \qquad -S_{A\cup B}-S_{B\cup C} - S_{C\cup A}
\label{eq:tmi}
\end{align}
evaluated for three consecutive intervals $A$, $B$, $C$ of length $L/4$.
$\tmi$ as defined above vanishes in area-law entangled states, has an extensive (negative) value in the volume-law phase, and is finite at critical points.
This makes it particularly useful in estimating the location of critical points~\cite{Gullans2019A}, as it gives rise to crossings with very limited finite-size drift (the entanglement entropy, on the other hand, has a logarithmic drift at criticality which makes finite-size scaling harder).
The single-parameter scaling ansatz
\begin{equation}
\tmi (q, L) \sim F[(q-q_c)L^{1/\nu}] \;,
\label{eq:I3scaling}
\end{equation}
where $q$ parametrizes the measurement ensemble,
can be used to estimate the correlation length critical exponent $\nu$.

\begin{figure}
\centering
\includegraphics[width=\columnwidth]{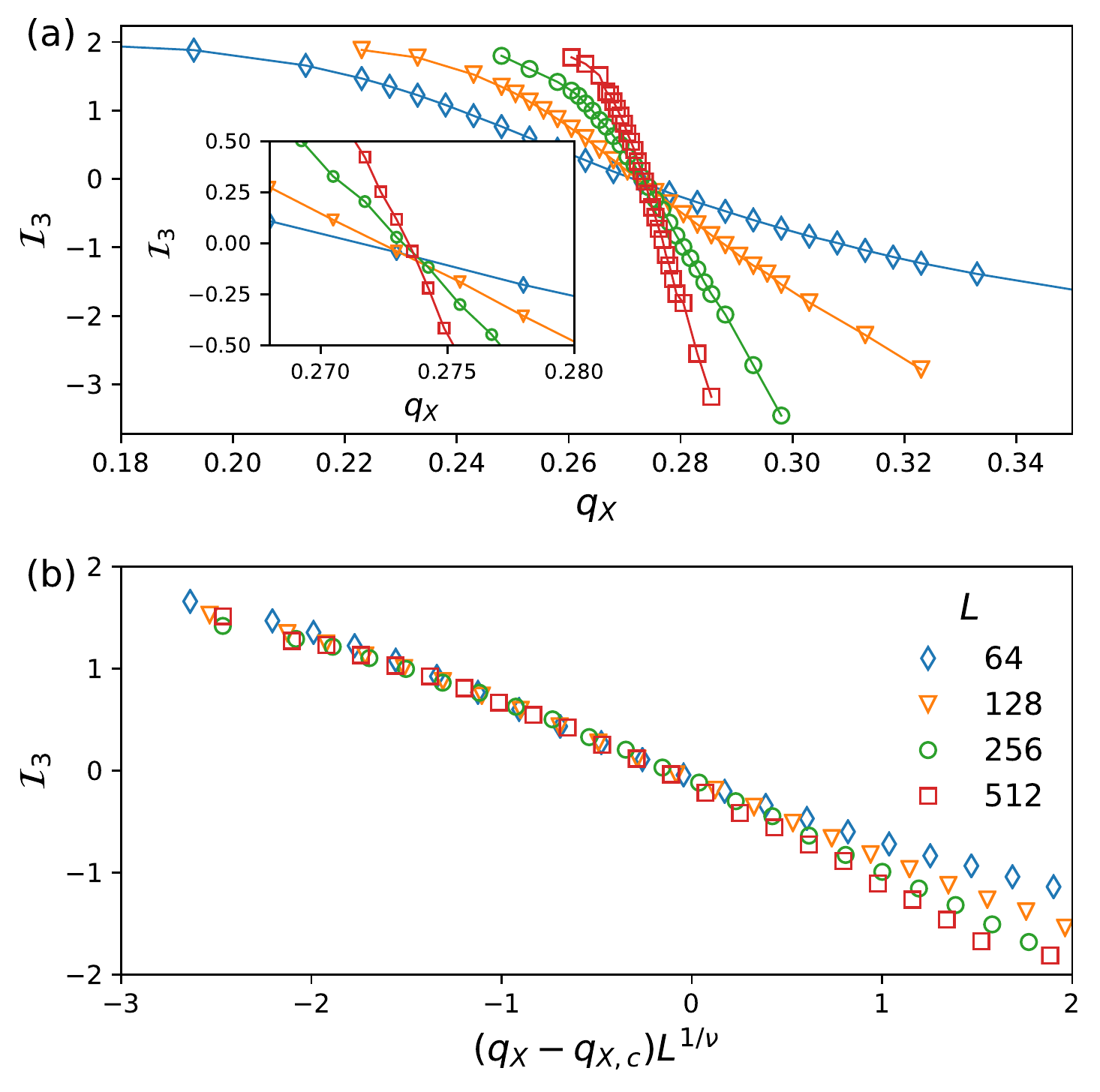}
\caption{Entanglement transition in the factorizable ensemble $q = (0,q_X,1-q_X,0)$ with range $r=3$. Data obtained from numerical simulations with the stabilizer method, averaged over $10^2$ to $10^4$ realizations of the random dynamics (depending on $L$).
(a) Tripartite mutual information $\tmi$ as a function of $q_X$.
The value $\tmi = 2$ on the area-law side is due to the presence of two integrals of motion (bits of global entanglement).
Different sizes $64\leq L \leq 512$ show a crossing at $q_X = q_{X,c} = 0.274(2)$ (inset).
(b) Scaling collapse of the data with exponent $\nu = 1.1$. 
\label{fig:r3_transition}}
\end{figure}

We consider the $r=3$ factorizable models and focus for simplicity on the $q_Z=0$ line, where this model has an area-to-volume critical point (shown by the star in Fig.~\ref{fig:triangles}(a)) at $q_{X,c} = 0.274(2)$.
We note that this model, consisting of Pauli strings $\{XXX,XXY,\dots YYY\}$, has two independent `integrals of motion', or symmetries.
These are global Pauli strings which commute with all measurements\footnote{For a finite system with periodic boundary conditions, this is true only if $L$ is multiple of 3. We use open boundary conditions in this calculation.}: $\prod_j Z_{3j} Z_{3j+1}$ and $\prod_j Z_{3j+1} Z_{3j+2}$.
These operators contribute two bits of \emph{positive} tripartite mutual information $\tmi$. 
This offsets the value of $\tmi$ in the area-law phase to $\tmi=2$, as seen in Fig.~\ref{fig:r3_transition}(a).
In the vicinity of the critical point, the scaling ansatz Eq.~\eqref{eq:I3scaling} yields a correlation length exponent $\nu = 1.1(1)$, although substantial corrections to the finite-size scaling remain visible on the volume-law side, as shown in Fig.~\ref{fig:r3_transition}(b). 
This is due to the low entropy density of the volume-law phase in this model.
Additionally, we find that the entanglement entropy at the critical point obeys $S(\ell) \simeq K \ln \ell$ with $K = 1.0(1)$.

We also study the local order parameter introduced in Ref.~[\onlinecite{Gullans2019B}], i.e. the long-time limit of the entanglement $S_R(t) \equiv S(\rho_R(t))$ of a reference qubit $R$ initialized in a Bell pair state with a qubit at position $x$ in the system. 
In the area-law (pure) phase $S_R$ vanishes as the reference is quickly disentangled,  while in the volume-law (mixed) phase entanglement persists for exponentially long times. 
At criticality $S_R$ vanishes parametrically slowly in system size, $S_R(t) \sim G(t/L^z)$ for some function $G$.
We find a dynamical exponent $z=1$, consistent with the transition being described by a CFT, in agreement with previous studies on the transition in unitary-projective circuits.
More specific evidence of a CFT description for one of these critical models (the $r=2$ factorizable MOM) is presented in Sec.~\ref{sec:code}, where we study time-dependent QECC properties.

Additionally, in Appendix~\ref{app:mom_lbit} we examine in a similar way a different family of MOMs, based on the $l$-bit unitary-projective circuit from Sec.~\ref{sec:lbit}. There too we find an area-to-volume critical point with a correlation length critical exponent $\nu = 1.1(1)$ and dynamical exponent $z=1$. The coefficient of the logarithmic divergence in the entropy is $K=0.8(1)$. 

The critical properties of these two examples are compatible, pointing to the possibility of a unique universality class for measurement-only entanglement transitions in 1D.
Additionally the correlation length exponent found here ($\nu = 1.1(1)$) is lower than the one found for the EPT in hybrid Clifford circuits~\cite{Gullans2019A} ($\nu = 1.28(2)$), suggesting that the MOM universality class may be distinct from the Clifford unitary-projective one.
However, the limited resolution on critical exponents and the large variety of other models we have not studied mean that these results should be viewed as only a preliminary investigation of these critical points, and that more thorough investigations are needed to settle this issue.


\section{Bipartite ensembles and Quantum Order\label{sec:bipartite}}

The results of Sec.~\ref{sec:ensembles} show that a volume-law phase is the generic outcome for ``long enough'' and ``random enough'' Pauli strings -- but making these qualifiers more specific remains challenging.
In light of this, rather than focusing on what enables a volume-law phase, one can take the opposite view of searching for \emph{obstructions} to this generic outcome.
In this Section we discuss a class of models, characterized by a special algebraic property -- \emph{bipartition} of the frustration graph -- that present one such obstruction, and give rise to qualitatively different phenomenology, including the possibility of steady states characterized by different types of quantum order. 
Whether other exceptions such as this one exist is an interesting question for future research.
 
\subsection{Absence of volume-law phase}

\begin{figure}
\centering
\includegraphics[width=\columnwidth]{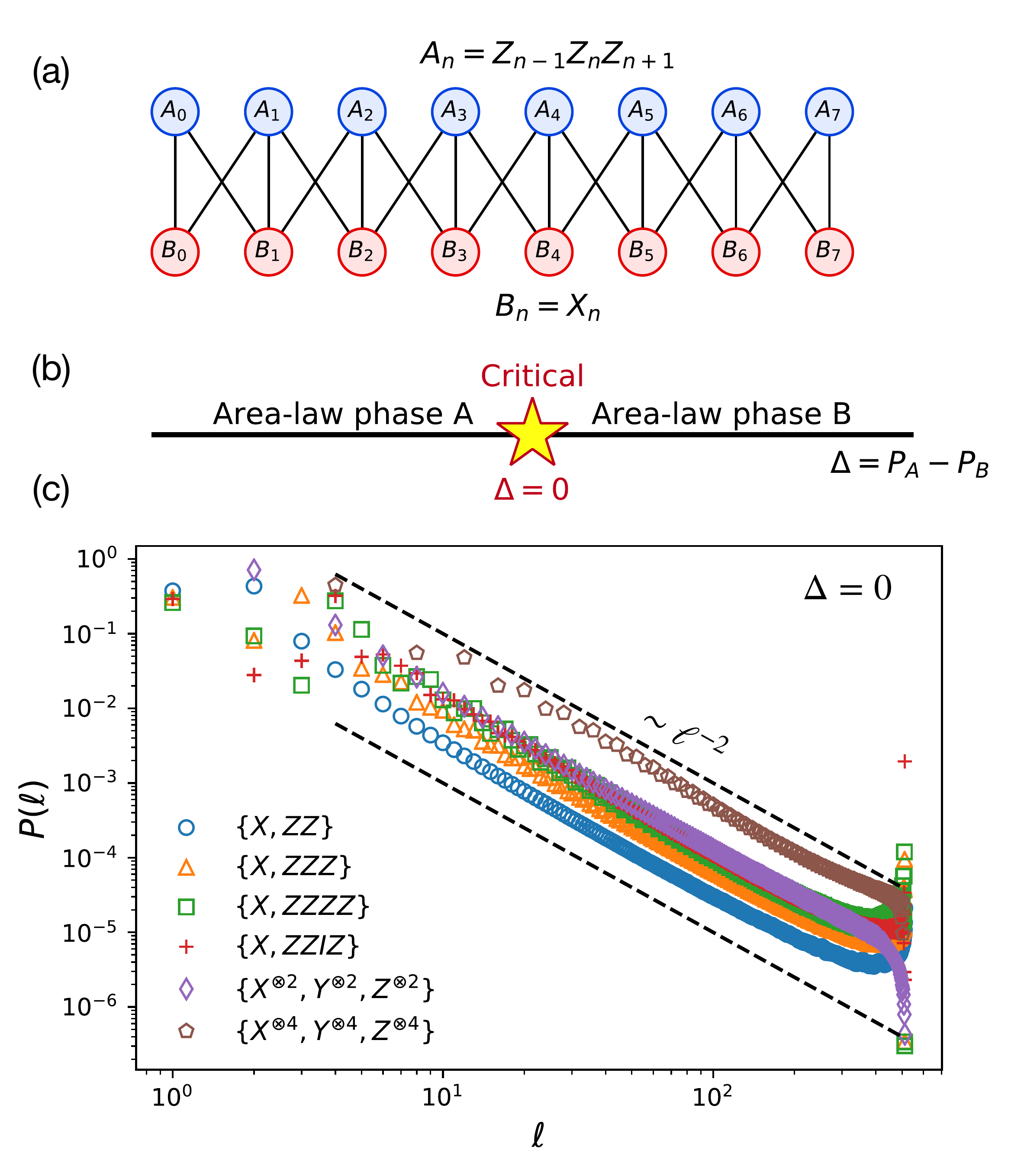}
\caption{\label{fig:pofl}
(a) Example of a frustration graph for a bipartite ensemble, $\{X,ZZZ\}$. Vertices represent operators and edges represent anticommutation.
(b) General phase diagram of 1D bipartite ensembles: as a function of $\Delta = P_A-P_B$, there are two area-law phases separated by a self-dual critical point at $\Delta=0$.
(c) Probability distribution of stabilizer length $P(\ell)$ in several bipartite ensembles at the $\Delta = 0$ critical point in a system of $L=512$ qubits (data aggregated from $10^3$ runs for each ensemble). 
A power-law tail $P(\ell) \sim \ell^{-2}$, corresponding to logarithmic entanglement, is seen in all models.}
\end{figure}

We consider ensembles with only two species of Pauli strings, $A$ and $B$, whose intra-species commutation relations are trivial: $[A_i, A_j] = [B_i, B_j] = 0$. 
This means the frustration graph is bipartite, i.e. $A$-type vertices are only connected to $B$-type vertices, and vice versa (Fig.~\ref{fig:pofl}(a)). 
Physically, this situation can describe two quantum error correcting codes whose stabilizers are $\{A_i\}$ and $\{B_i\}$ respectively.
These are mutually incompatible -- stabilizers for the $A$ code are interpreted as errors by the $B$ code and vice versa.
Both types of operators are measured concurrently with rates proportional to the probabilities $P_A$, $P_B$.
The phase diagram is thus one-dimensional, parametrized by the bias $\Delta = P_A-P_B \in [-1,1]$.

To characterize this phase diagram it is helpful to use the frustration tensor $\Gamma^{\alpha\beta}_\ell$ introduced in Sec.~\ref{sec:frustration}.
The only nontrivial sector in the frustration tensor is
 $\Gamma^{AB}_\ell =  \Gamma^{BA}_{-\ell} \equiv \gamma_\ell$.
A spatial reflection $\ell \mapsto -\ell$ implements a species duality transformation $(A,B) \mapsto (B,A)$, $\Delta \mapsto -\Delta$.
Since spatial reflection cannot change the entanglement phase, the phase diagram must be symmetric about $\Delta = 0$.
The extrema $\Delta = \pm 1$ are fully un-frustrated -- only one operator species is measured and the system is in the simultaneous eigenstate of all $\{A_i\}$ (or $\{B_i\}$) operators, which is area-law entangled as long as these are local. 

The question, then, is what the interior of the phase diagram $-1<\Delta<1$ looks like.
Possibilities include (i) an intervening volume-law phase in an interval $|\Delta|<\Delta_c$, (ii) an extended critical region, (iii) a single critical point at $\Delta = 0$.

Numerical simulations of all bipartite graphs with range $r \leq 6$ in one dimension show that the answer is always (iii): a single, isolated critical point at $\Delta = 0$, surrounded by area-law phases, see Fig.~\ref{fig:pofl}(b). 
Before addressing the critical points, we emphasize that this means the \emph{absence} of volume-law phases in these models, in sharp contrast with the general phenomenology discussed in Sec.~\ref{sec:ensembles}. 
Weakly perturbing the models to break the bipartition generally opens up a volume-law phase near the critical point.

Based on this numerical evidence, we conjecture that bipartition of the frustration graph poses a general obstruction to the existence of volume-law phases. 
This conjecture is further corroborated by recent results on two-dimensional MOMs~\cite{Lavasani2020B}. 
The purification dynamics of an initially fully mixed state in these bipartite MOMs can always be expressed in a gauge where each stabilizer generator is either a product of $A$ operators only or of $B$ operators only, with no mixing. 
This corresponds to two \emph{classical} error-correcting codes. 
It is possible that a QECC with this structure may be too weak to sustain a mixed phase; proving this would be an interesting goal for future work. 

An intuitive picture for these critical points goes as follows. At maximum bias $\Delta=1$ ($P_B = 0$) the dynamics is fully un-frustrated and projects the state into the $A$ code space (with area-law entanglement). As infrequent $B$ measurements are introduced ($0<\Delta<1$) small patches of $B$ code (i.e. intervals in space where $B_x\ket{\psi} = \ket{\psi}$) are constantly created and destroyed over a background of $A$ code, and are prevented from spreading beyond a finite length scale by the frequent $A$ measurements. The same, with $A\leftrightarrow B$, is true at $\Delta<0$. At $\Delta = 0$, however, neither code dominates and the formation of long stabilizers becomes possible. 

In Fig.~\ref{fig:pofl}(c), we show the stabilizer length distribution $P(\ell)$ in the ``clipped gauge''~\cite{Li2018} for various bipartite models at the critical point $\Delta=0$; all of them exhibit a power-law tail $P(\ell) \sim K\ell^{-2}$.
The coefficient $K$ is found to increase with the range $r$ of the bipartite ensembles.
This coefficient is related to the entropy via $S(\ell) \sim \frac{K}{2}\ln(\ell)$ (one bit of entropy is carried by \emph{two} stabilizers straddling a boundary, hence the factor of $\frac{1}{2}$). 
The different values of $K$ suggest that these critical points are described by different critical theories. 
If so, this class of models would introduce a wide class of novel entanglement critical points whose position is exactly known and fixed by a duality ($A \leftrightarrow B$), unlike e.g. hybrid circuits where $p_c$ must be determined numerically.
This could be a useful setting for future studies of the underlying critical theory.

\subsection{Ordered area-law phases}

This class of critical points in bipartite MOMs is also particularly interesting when viewed as a dynamical phase transition between different species of area-law states with distinct patterns of quantum order, e.g. the $A$ ($B$) operators may be the stabilizers of a trivial (topological) phase. 
This is similar in spirit to transitions between different MBL phases with characteristically different l-bits~\cite{lpqo}. 
Exactly solvable models with extensively many local commuting projectors often describe renormalization group fixed points for different phases (or $l$-bit representations of MBL phases); the eigenstates of such models are also simultaneous eigenstates of all the local projectors and can display non-trivial quantum order. 
Drawing the $A$ and $B$ operators from the sets of projectors characterizing two different phases can yield late-time steady states with different patters of order~\footnote{See also Refs.~[\onlinecite{Lavasani2020}] and [\onlinecite{Hsieh2020}] for related contemporaneous work on ordered phases in unitary-projective and measurement-only dynamics}. 
For example, $\{A_i = X_i\}$ and $\{ B_i = Z_{i-1}X_iZ_{i+1}\}$ correspond to the trivial and symmetry-protected-topological (SPT) paramagnet respectively. Likewise, we could pick one or both of $A$ and $B$ to be the stabilizers of a topological code such as the toric code, in which case the steady states in the area-law phase would display non-trivial (and non-local) order characteristic of the topological phase.
In 1D, any choice of bipartite frustration graph, $\Gamma^{AB}_\ell \equiv \gamma_\ell$ (realized e.g. by $\{A_i = X_i\}$, $\{B_i = \prod_\ell Z_{i+\ell}^{\gamma_\ell}\}$), yields a trivial phase and a phase with up to $r-1$ distinct $\mathbb{Z}_2$ symmetries, whose operators are encoded in the kernel (over $\mathbb{Z}_2$) of the banded matrix $M_{ij} \equiv \gamma_{i-j}$.

\begin{figure}
    \centering
    \includegraphics[width=0.99\columnwidth]{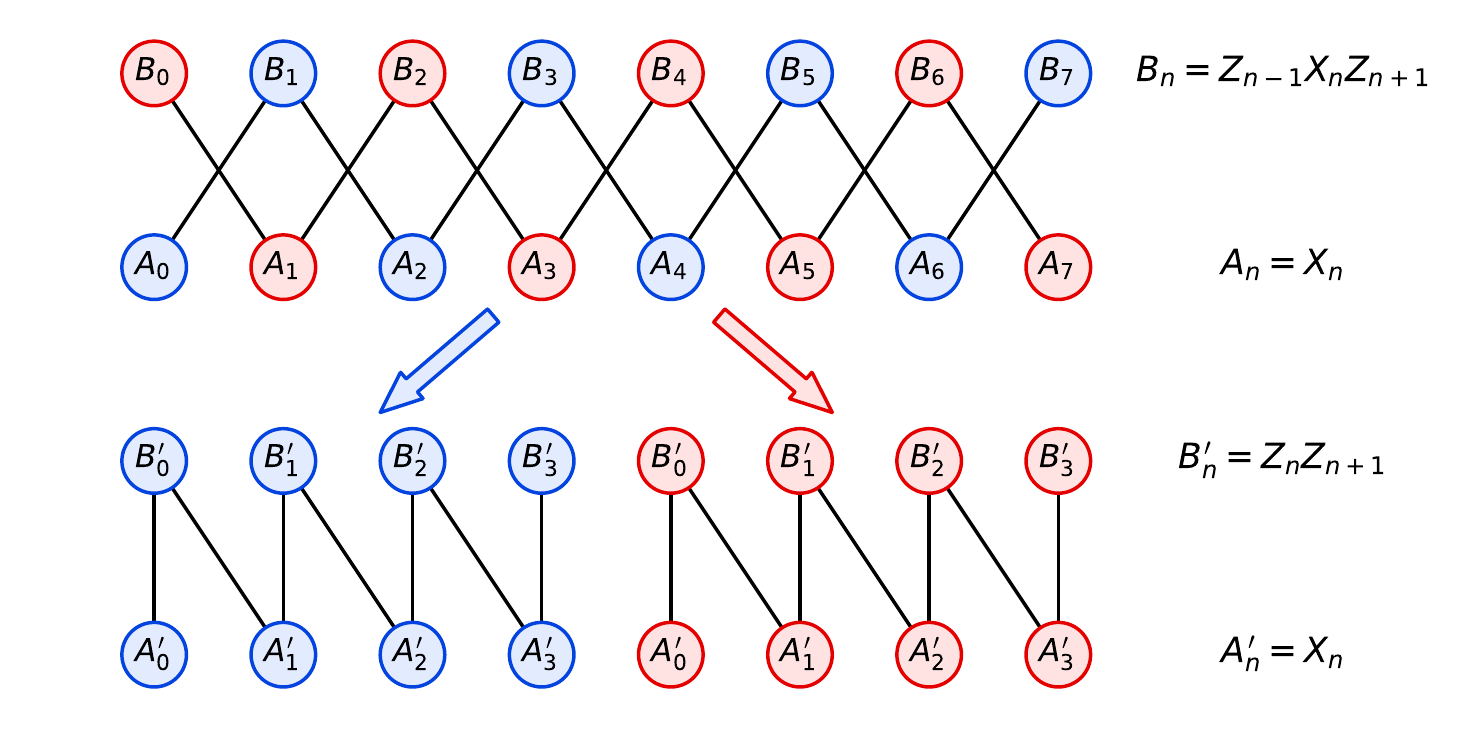}
    \caption{The frustration graph for the $\{X,ZXZ\}$ MOM (top) splits into two disconnected subgraphs that correspond to the $\{X,ZZ\}$ MOM (bottom). 
    The two models have identical purification phase diagrams and critical properties, up to a factor of 2 in the entanglement entropy.
    }
    \label{fig:decomposition}
\end{figure}

Furthermore, the frustration graph can reveal equivalences between seemingly different transitions. For example, Ref.~[\onlinecite{Skinner2019}] considered a free-fermion MOM where operators were drawn from the sets $\{Z_i Z_{i+1}\}, \{X_i\}$. This is a bipartite MOM with a transition between a trivial area-law phase and one with $\mathbb{Z}_2$ order that is understood as loop percolation. 
The frustration graph reveals that the ensemble $\{A_i=X_i\}$ and $\{B_i = Z_{i-1}X_iZ_{i+1}\}$ later studied in Ref.~[\onlinecite{Lavasani2020}] is actually equivalent to two decoupled copies of the $\{A_i = X_i\}$ and $\{B_i = Z_{i-1}Z_{i}\}$ ensemble, Fig.~\ref{fig:decomposition}.  
This implies the two models have identical purification phase diagrams and critical exponents, and the coefficients $K$ in $S\sim K\ln(\ell)$ for the two models at criticality are related by a factor of $2$. The equivalence between these two examples is simple to establish using our frustration graph formalism and hints at the predictive powers of this formalism in classifying the phase structure of  MOMs.
Additional examples of equivalence between MOMs are discussed in App.~\ref{app:f_equivalence}.

\begin{figure}
\centering
\includegraphics[width=0.99\columnwidth]{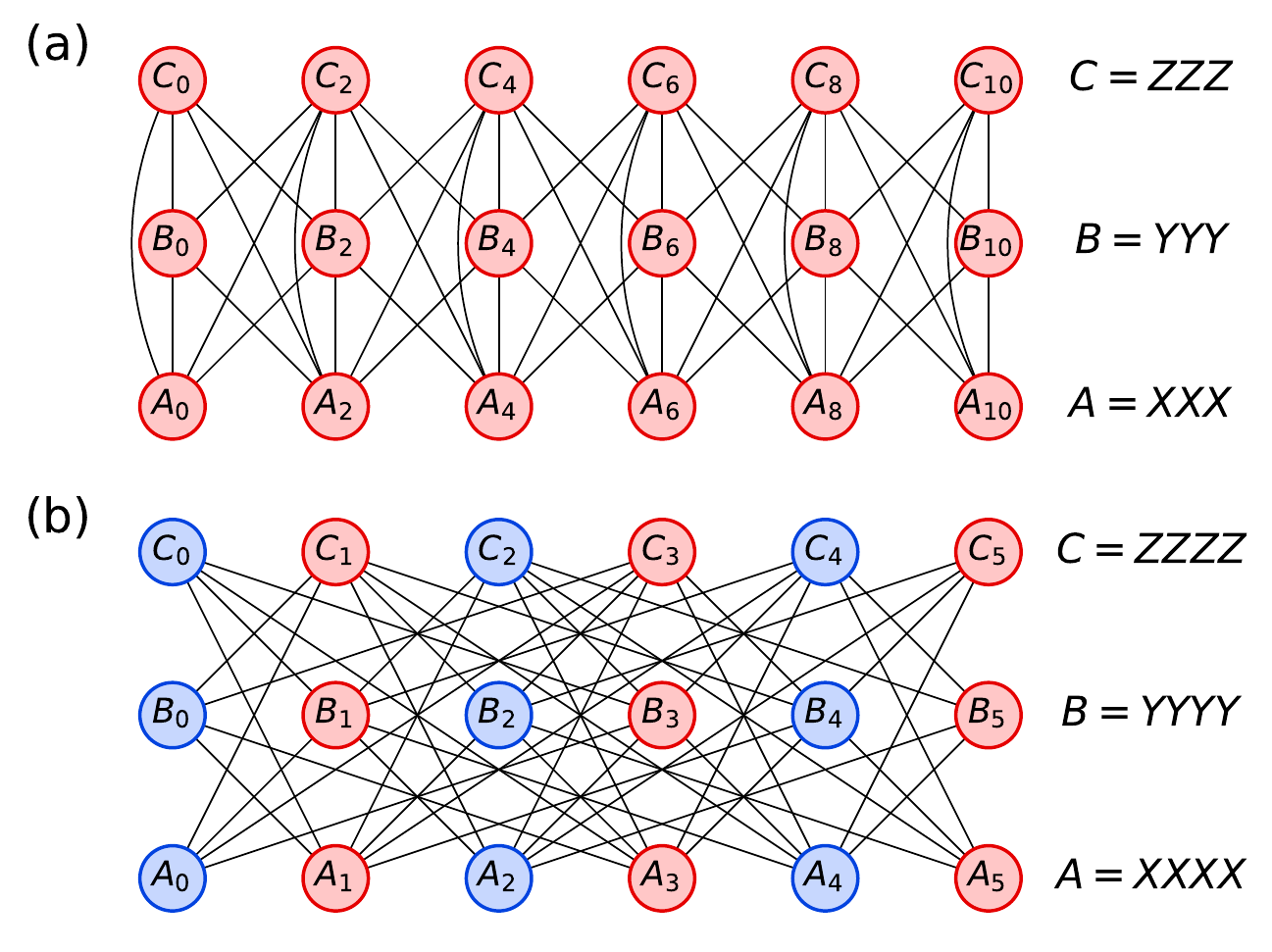}
\caption{\label{fig:evenodd}
Frustration graphs of the $\{A=X^{\otimes r}, B=Y^{\otimes r}, C=Z^{\otimes r}\}$ ensembles with (a) $r=3$ and (b) $r=4$, for a finite system. 
Each vertex represents one operator; edges connect anticommuting operators.
For odd $r$, the graph decomposes into two identical, disconnected subgraphs -- (a) shows only one of them. 
For even $r$, the graph is connected but bipartite (as indicated by the color scheme in (b)), and there is no volume-law phase despite the higher connectivity of the graph.}
\end{figure}

We conclude this Section by noting that a bipartition in the frustration graph need not be between operator species (as defined above).
The ensembles $\{A = X^{\otimes r} , B = Y^{\otimes r}, C = Z^{\otimes r}\}$, for example, exhibit strikingly different behavior depending on whether $r$ is even or odd -- a volume-law phase is possible for odd $r$ ($r>1$), but not for even $r$, which is area-law or at most critical (depending on the probabilities $P_{A,B,C}$). 
As we have seen, the odd-$r$ behavior (which admits a volume-law phase) is the generic one.
The reason for the anomalous behavior at even $r$ is that the frustration graph is bipartite \emph{spatially}: all strings starting on even sites $\{A_{2j}, B_{2j}, C_{2j}\}$ commute amongst themselves; the same holds for those starting on odd sites, $\{A_{2j+1}, B_{2j+1}, C_{2j+1}\}$, see Fig.~\ref{fig:evenodd}(a).
This is not true for odd $r$, where e.g. $A_0$, $B_0$ and $C_0$ anticommute pairwise and thus form a triangular subgraph, see Fig.~\ref{fig:evenodd}(b).

\section{Quantum code properties\label{sec:code}}

The measurement-only dynamics induced by an ensemble of observables $\{ O_{\alpha} \}$ in the volume-law phase generates a random quantum code that protects information {\it against the operators $\{ O_\alpha\}$ themselves}. In this Section we examine the properties of these dynamically generated quantum error-correcting codes~\cite{Choi2019b,Gullans2019A,Fan2020}.

A stabilizer quantum error-correcting code is conventionally labelled by a triplet of integers $[[n,k,d]]$, where $n$ is the number of physical qubits, $k$ is the number of encoded logical qubits, and $d$ is the code distance.
The ratio $R = k/n$ is also known as the code rate. 
The code distance $d$ is the minimum weight (number of non-identity sites) of an undetectable logical error -- an operator $\mathcal E$ that commutes with the stabilizer group but does not belong to it.
Intuitively, larger $d$ means that more errors can be corrected.
For a family of $[[n,k,d]]$ codes to have a finite error correction threshold, the distance has to diverge in the thermodynamic limit $n \to \infty$.
There is generally a trade-off between code rate and distance, manifested e.g. in the `quantum Singleton bound'~\cite{Knill2000}, $\frac{k}{n}+2\frac{d-1}{n} \leq 1$.

In the present case, we have $n=L$ (number of physical qubits) while $k$ is the entropy of the steady state of the measurement-only dynamics. 
The computation of $d$ is thought to be exponentially hard in $L$ in general. 
Here we consider a related quantity that can be computed in time $\text{poly}(L)$ for stabilizer states: the \emph{contiguous} code distance, defined as \cite{Bravyi09,Gullans2019A}
\begin{equation}
\ell_x = \min \{|A_x|:\ \exists \ {\mathcal E} \text{ supported in } A_x\}
\label{eq:contiguous_def}
\end{equation}
where $A_x$ is a contiguous interval of the chain containing site $x$ and $\mathcal E$ is a logical operator, as defined above. This satisfies $d\leq \ell_x$: if there exists a logical operator supported in $A_x$, then its weight is at most $|A_x| = \ell_x$.
A related quantity that was used in a no-go theorem for self-correcting memories in two dimensions is the linear code distance~\cite{Bravyi09} $\ell_{\min} = \min_x \ell_x$. 
In the following we will consider the \emph{averaged} contiguous distance~\cite{Gullans2019A} $\langle \ell \rangle \equiv \frac{1}{L} \sum_x \ell_x$, since the dynamics generating the code is (statistically) invariant under spatial translations; results for $\ell_{\min}$ are qualitatively similar.
We define the contiguous code distance of pure stabilizer states (i.e., an $[[n,0]]$ code) as zero.  
Thus $\langle \ell \rangle$, with this convention, equals the probability that the system is in a mixed state (and thus defines a code) times the averaged contiguous distance of those realizations.
For simplicity in the following we use `distance' to mean `averaged contiguous code distance', and we use the notation $\langle \ell\rangle$ to denote averaging over both space and realizations of the dynamics.

As we mentioned earlier, a MOM with measurement ensemble $\mathcal{E} = \{O_\alpha\}$ in the mixed phase generates a quantum code that must necessarily detect all elements of $\mathcal{E}$ as errors (up to exponentially rare events). To see why, let us imagine that an operator $O_{\alpha,i} \in \mathcal{E}$ had a finite probability $p$ of being an undetectable logical error for a steady-state code $\rho$; measuring such operator would partially purify the state, giving an expected change in entropy $\overline{\delta S(\rho)} \leq -pP_\alpha/L$ over the following time step; but since $\rho$ is a steady-state code, its entropy must decay exponentially slowly, hence $p$ must be exponentially small in $L$. Typical steady-state codes can thus detect all elements of $\mathcal E$ as errors. Given this fact, a natural question to ask is how ``specialized'' these codes are, i.e. whether they can also detect arbitrary errors (up to some distance) beyond those in the ensemble $\mathcal{E}$ that defines the dynamics.

\begin{figure}
    \centering
    \includegraphics[width=\columnwidth]{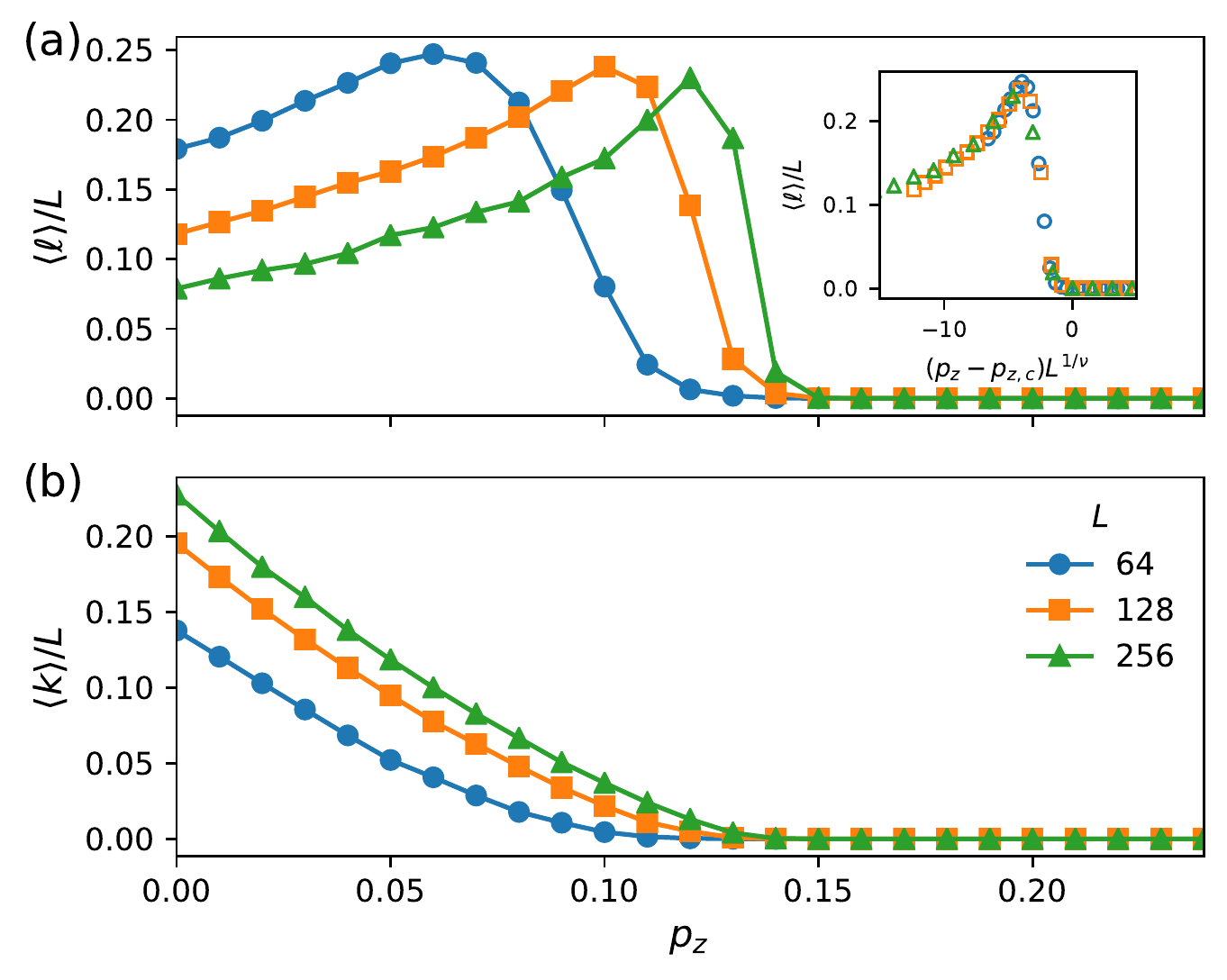}
    \caption{Quantum code properties of the $l$-bit ensemble (App.~\ref{app:mom_lbit}) with range $n=4$ and varying probability of $Z$ measurements $p_z$ (data taken at time $t=4L$), averaged over at least 100 realizations. (a) Contiguous code distance, computed as described in the text. The scaling is subextensive deep in the mixed phase and extensive near criticality. Inset: scaling collapse with $p_{z,c} = 0.15$ and $\nu = 1.1$. (b) Code rate. 
    Both the rate and the distance evolve smoothly from the $p_z = 0$ point.}
    \label{fig:lbit_code}
\end{figure}

To gain some insight into this problem we consider an MBL-inspired MOM (see App.~\ref{app:mom_lbit}), with the addition of single-site $Z$ measurements. From the phase diagram of the related hybrid circuit model, Fig.~\ref{fig:lbitphasediag}, we know that a volume-law phase is possible at sufficiently low $p_z$. The steady-state code necessarily detects single-qubit $Z$ errors at $p_z>0$ (within the mixed phase), as we discussed above.
Does this remain true for $p_z = 0$?
In other words, does the code become vulnerable to single-qubit $Z$ errors if these are not explicitly injected in the dynamics? 
To address this question we simulate the $l$-bit MOM with range $n=4$ and variable $p_z$: 
with probability $p_z$ we measure a single-site $Z$, otherwise we measure (with uniform probability) one of the $2^{2n-1}$ Pauli strings in Eq.~\eqref{eq:lbit_momensemble}.
We find no singular change in either $\langle k \rangle$ or $\langle \ell\rangle$ as $p_z$ is turned on, as shown in Fig.~\ref{fig:lbit_code}.
The only effect of $p_z$ is to move the dynamics closer to the transition (and eventually into the area-law phase), which as expected increases the distance $\langle \ell\rangle$ at the expense of the rate $\langle k \rangle/L$. This happens smoothly in $p_z$.
The behavior of $\langle \ell \rangle$ in this model is similar to that observed in hybrid circuits~\cite{Gullans2019A, LiFisher2020}, with a subextensive scaling $\langle \ell \rangle \sim L^a$ ($0<a<1$) deep in the mixed phase, an extensive scaling near the critical point, and a drop to zero in the pure phase (the latter is due to the vanishing probability of the state remaining mixed and defining a code: we have set $\ell=0$ for pure states).

\begin{figure}
    \centering
    \includegraphics[width=\columnwidth]{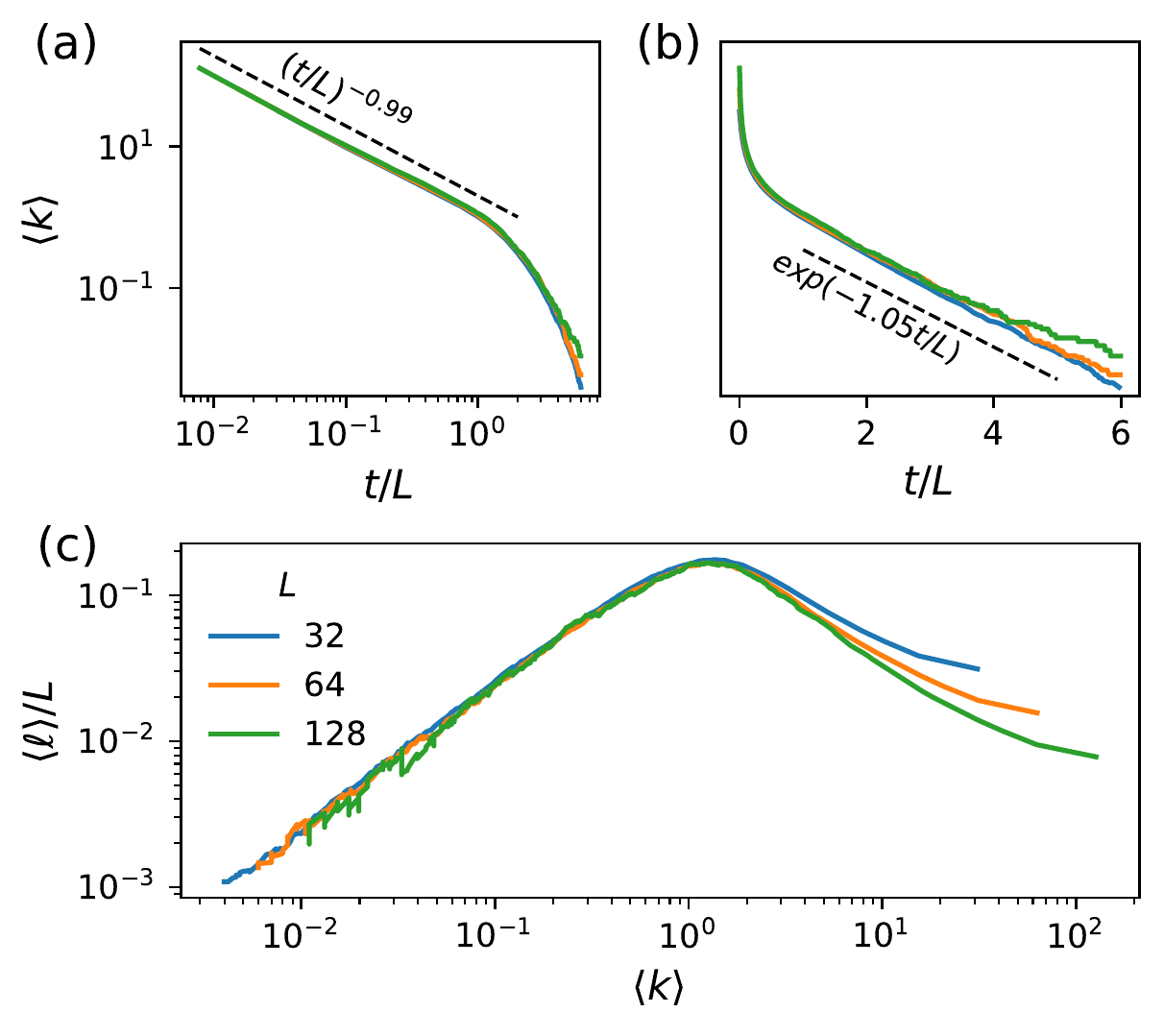}
    \caption{Quantum code properties of a critical MOM (the $r=2$ factorizable model of Fig.~\ref{fig:triangles}(b) at the $\mathbf q = \frac{1}{3}(1,1,1)$ point), averaged over $10^3$ realizations of the dynamics. 
    (a,b) Average number of encoded qubits, $\langle k \rangle$, as a function of time starting from the fully mixed state. 
    $\langle k \rangle$ depends only on $t/L$ (dynamical exponent $z=1$), decays as $L/t$ at early times (a) and exponentially at late times (b).
    (c) Average contiguous distance $\langle \ell \rangle$ as a function of $\langle k \rangle$ (time progresses right to left). 
    The distance peaks (and becomes extensive) when $\langle k \rangle \simeq 1$.
    }
    \label{fig:critical_code}
\end{figure}

We conclude this Section by moving from the volume law phase to critical points, where the system eventually purifies and thus does not form a quantum code in the steady state. 
However, the parametrically long timescale for purification allows us to probe the time-dependent code properties of the mixed state as it gradually purifies. 
To be concrete, we consider the $r=2$ factorizable MOM of Fig.~\ref{fig:triangles}(b) at the central point, $\mathbf{q}_0 = \frac{1}{3}(1,1,1)$, i.e. we measure the 9 Pauli strings $\sigma_a\otimes \sigma_b$, $a,b\in\{X,Y,Z\}$, with equal probability. This model is in the middle of a critical phase.
In Fig.~\ref{fig:critical_code}(a,b) we show the decay of the average number of encoded qubits $\langle k\rangle$. We find that $\langle k\rangle$ depends on time only through the ratio $t/L$, in agreement with the dynamical exponent $z=1$.
The decay is consistent with $\langle k\rangle \sim L/t$ at early times ($t\lesssim L$), then crosses over to exponential, $\langle k\rangle\sim e^{-ct/L}$.
This behavior was indicated as evidence of an underlying $(1+1)$-dimensional CFT: it corresponds to a one-parameter dependence on the `cross ratio' $\eta$ computed from the endpoints of the entanglement cuts~\cite{Li2020}.
The behavior of the distance $\langle \ell \rangle$ during the dynamics is also interesting (see Fig.~\ref{fig:critical_code}(c)). 
We start from $\langle \ell \rangle = 1$ at $t=0$ (as all Pauli strings, including single-site ones, are logical operators for the fully mixed state).
Then, as $\langle k \rangle$ decays, the distance increases.
This lasts until $t\simeq L$ and $\langle k \rangle\simeq 1$ (one logical qubit left in the system), where the distance is extensive.
After that, $0\leq \langle k \rangle < 1$ essentially represents the probability that the state is mixed.
Consequently the distance decays as $\langle \ell \rangle \sim L\langle k\rangle \sim Le^{-ct/L}$.

\section{Locality and information spreading \label{sec:loc}}

The effect of local measurements on entangled states was famously described as a ``spooky action at a distance''~\cite{Einstein1935, EinsteinLetter}.
In these models, where the entirety of the dynamics is made up of measurements on entangled states, there is good reason to expect spooky surprises. 
For instance, unlike local unitary circuits which have a strict light-cone, local projective measurements allow for the creation of arbitrary-range entanglement on an $O(1)$ timescale, using two layers of local measurements  acting on a product state. 
One can see this as follows~\cite{Zeilinger1993}:
Start from a $Z$-product state on a chain of length $L$, with stabiizer generators $\{ g_i = Z_i:\ i=1,\dots L\}$.
Measure the two-site operators $X_1 X_2, X_2 X_3, \dots X_{L-1} X_L$ (all commuting, and thus measurable at the same time). This creates one bit of mutual information between sites $(1,L)$, represented by the stabilizer $g = Z_1\cdots Z_L$. 
Then, measuring operators $Z_2, \dots Z_{L-1}$ (again all commuting) leaves sites $(1,L)$ in a Bell pair state.
While fine-tuned (and thus unlikely to occur if the measurements are placed randomly in spacetime), this example shows that there is in general no strict light cone for the production of entanglement or correlations in this type of dynamics~\cite{Li2020}. 
Any emergent light cone must be statistical in nature -- i.e., must reflect the fact that histories that produce entanglement outside the putative light cone are possible but rare. 

The propagation of information in quantum systems is described by the spreading of local operators evolved in the Heisenberg picture~\cite{Maldacena2016,Swingle2016,Nahum2018,VonKeyserlingk2018}.
In the presence of measurements, the Heisenberg picture is problematic, since the Born probabilities needed to choose projectors must be computed on a state\footnote{This is not an issue when averaging over measurement outcomes, in which case the adjoint of the quantum channel describing the (mixed) state evolution is perfectly well defined. Operator spreading in open systems was previously studied in Refs.~[\onlinecite{otocdiss1, otocdiss2,otocdiss3,otocdiss4}].}.
Nevertheless, one can still ask how information spreads across the system. In what follows we propose a diagnostic for information spreading and verify the emergence of a statistical light-cone.

\begin{figure*}
\centering
\includegraphics[width=\textwidth]{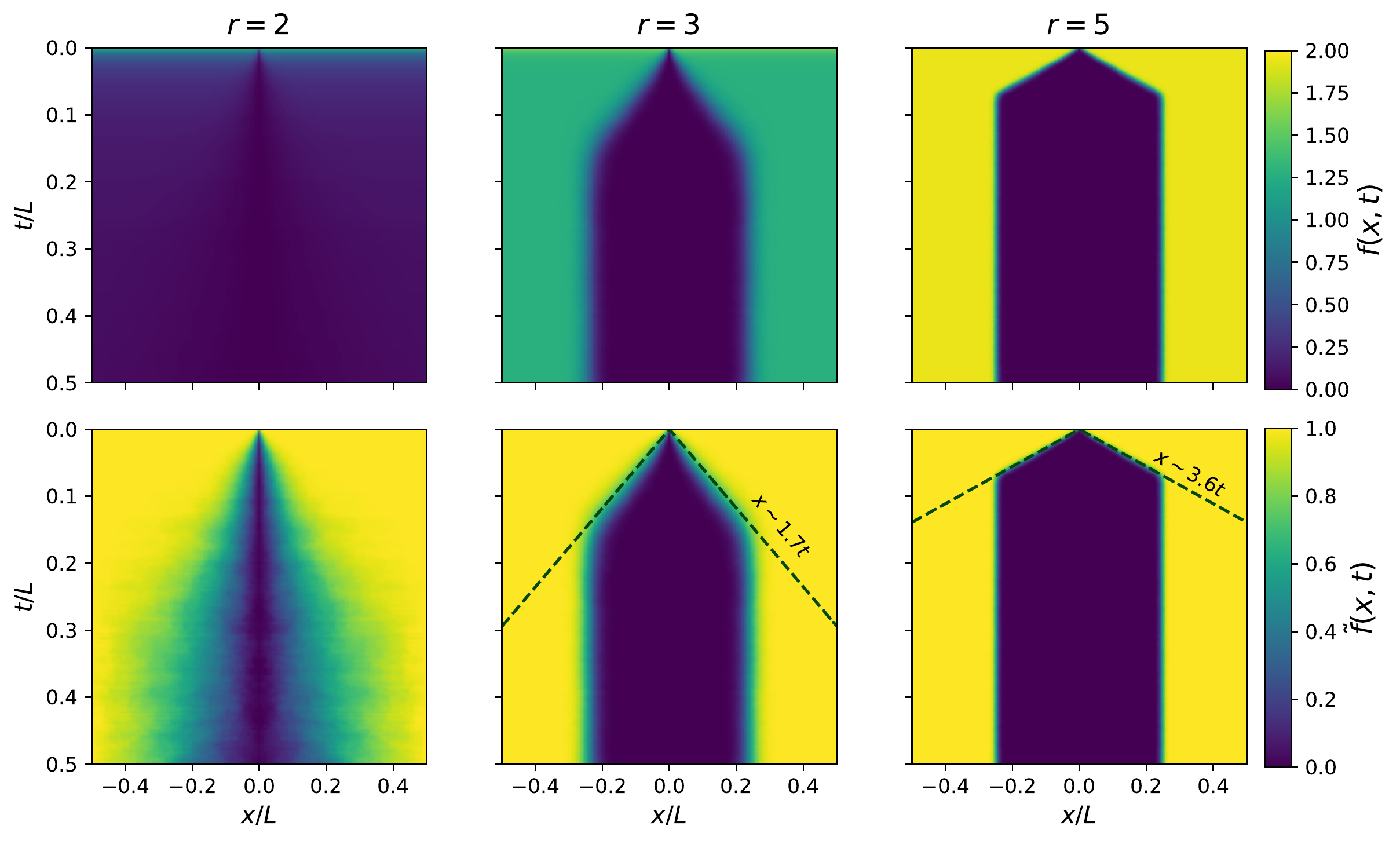}
\caption{Information spreading in measurement-only dynamics.
Top: $f(x,t)$ as defined in Eq.~\eqref{eq:mutual_info} (mutual information between a reference initially entangled at $x=0$ and a region $[-x,x]$) for factorizable ensembles $q = \frac{1}{3}(0,1,1,1)$ of range $r=2$ (critical), 3 and 5 (volume-law) for a system of $L=255$ qubits. 
There is a clear ballistic light cone in the volume-law phase. The light cone saturates to half the system size. 
Bottom: the normalized quantity $\tilde{f}(x,t) = f(x,t)/2S_R(t)$ reveals that information spreading is bounded by a light cone in the critical phase as well. The approximate location of the wavefront is highlighted with a dashed line in the volume-law examples. The butterfly velocity $x\sim v_B t$ increases with increasing range $r$.
\label{fig:lightcones}}
\end{figure*}

For concreteness we focus on Clifford circuits in what follows, but the ideas are straightforward to generalize. Consider entangling a reference qubit $R$ at time $t=0$ to the center of a 1D chain of odd length $L=2l+1$ (with qubits numbered by $-l\leq n\leq l$), and subsequently running the measurement-only dynamics on the system.
Initially $R$ is in a Bell-pair state with qubit $n=0$.
After time $t$, $R$ may or may not be entangled with the system. The entanglement between $R$ and the system has previously been studied as an order parameter for the volume law phase~\cite{Gullans2019B}:
in the volume law phase, $R$ stays entangled with finite probability out to very long times. Crucially, if one assumes that there is a light-cone, $R$ is entangled only with some interval of the system, $[-x,x]$. 
The size of this subsystem is what captures information spreading in this setting.
One can estimate this by calculating the mutual information between $R$ and segments centered at the point of initial entanglement, $[-x,x]$, for variable $x$.
This defines a function on spacetime, 
\begin{equation}
f(x,t) = \mathcal I(R: [-x,x])|_t \;,
\label{eq:mutual_info}
\end{equation}
which quantifies how much information about the operators initially entangled with $R$ ($X_0$ and $Z_0$ at $t=0$) can be recovered by looking only at the region $[-x,x]$ at time $t$.
Equal-value contours of $f(x,t)$ thus capture the spread of these operators.

At the initial time, we have $f(x,0) = 2$ for all $x$: all segments $[-x,x]$ include the central qubit ($x=0$ is defined as containing the central qubit only). 
At late times, $f(L/2,t)$ is equal to the local order parameter $S_R(t)$ introduced in Ref.~[\onlinecite{Gullans2019B}], which has a finite value in the volume-law phase.
More precisely, in the volume law phase we expect $f(x,t \gg L)$ to approach zero for $x < l/2$ and a finite constant for $x>l/2$ (where $L = 2l+1$). 
This is because information hidden in a random state of $L$ qubits is recoverable with high probability from any subsystem of more than $L/2$ qubits (a result that follows from the quantum channel capacity of the erasure channel~\cite{Hayden2007, DiVincenzo1997}). 
At intermediate times, we expect $f$ to develop a ``hole'' near $x=0$ which progressively expands until eventually saturating to half the system.
This expectation is borne out by numerics on the factorizable ensembles (Sec.~\ref{sec:factorizable}) with $\mathbf{q} = (1,1,1)/3$ and ranges $r=3$ and 5, which are in the volume-law phase.
The information spreading is found to be ballistic, see Fig.~\ref{fig:lightcones}, with a `butterfly velocity' $v_B$ that increases with $r$. 
The final saturation value of $f(x,t \gg L)$ for $x>l/2$ is $2S_R(t)$, (twice) the total entanglement between system and reference, which is the order parameter for the volume-law phase and is a function of the model parameters, decreasing to zero as the transition to the area-law phase is approached.  

Outside the volume-law phase, $f(x,t)$ decays in time for all values of $x$.
It is nonetheless possible to define a normalized $\tilde{f}(x,t) \equiv f(x,t) / 2S_R(t)$, where again $2S_R(t) = f(L/2,t)$ is the mutual information between the reference $R$ and the whole system, and analyze the information spreading in the same way as for the volume law phase (though in practice this makes the data considerably noisier). 
Fig.~\ref{fig:lightcones} includes data for the same ensemble with range $r=2$, which is critical.
While the data is much noisier in this case (due to the majority of realizations becoming disentangled over short times and contributing no signal), the spread of information is still bounded by a finite velocity.  
The infinite entanglement velocity identified in Ref.~[\onlinecite{Li2020}] for measurement-induced critical points is not seen through this diagnostic. 
While nonlocal creation of entanglement through processes such as the `entanglement swapping' outlined above are likely happening, they seem to be statistically irrelevant to the dynamics of the encoded quantum information. For stabilizer circuits, the conditional trajectories where the reference qubit does not purify necessarily undergo purely unitary evolution, despite the nonunitary measurements.  Our results shown here indicate that this effective time-local random unitary evolution may also  have a spatially local description throughout the phase diagram.     

The application of this diagnostic to other models (both measurement-only and hybrid) is left for future work.


\section{Discussion \label{sec:discussion}}

In this work we have introduced a new type of entanglement phase transition, which arises in models whose dynamics consists entirely of local projective measurements. 
This is conceptually distinct from the transition in hybrid unitary-projective circuits, as the survival of the entangling phase is not due to the action of scrambling unitary gates, but rather by the measurements themselves, and more specifically by the \emph{frustration} of distinct types of measurements performed on the system.

In some cases (e.g. the truncated $l$-bit circuits we discussed in Sec.~\ref{sec:lbit}) these models can be equivalently viewed as either hybrid circuits or measurement-only models;
however, when viewed as hybrid circuits these models are extremely special, in that the entangling phase does \emph{not} exist at zero measurement rate. In other words, the unitary part of the circuit is fully non-scrambling, and thus volume-law entanglement is \emph{enabled} by the measurements.
This indicates that the conventional understanding of entanglement transitions in hybrid circuits as a competition between scrambling unitary dynamics (favoring a volume law phase) and projective measurement (favoring an area law phase) is incomplete;
measurements can play a more constructive role than previously thought.
One particularly striking consequence of this is that entanglement transitions are possible even in monitored MBL systems (see Fig.~\ref{fig:lbitphasediag}).

The class of models we have introduced is defined by many tunable parameters (as opposed to a single measurement rate or probability), giving rise to rich, multi-dimensional phase diagrams.
We found that the generic outcome for an ensemble of sufficiently long ($\gtrsim3$ sites), sufficiently random measurements is an entangling phase. Disentangling phases are possible for especially short or commuting observables, as one may expect; however, we have also uncovered more surprising exceptions. 
In Sec.~\ref{sec:bipartite} we showed that a special algebraic condition (bipartition of the frustration graph) seemingly prevents the formation of an entangling phase, even when measuring arbitrarily long Pauli strings. These models exhibit an interesting phenomenology, with dual area-law phases separated by a self-dual critical point, vastly generalizing a known example of this phenomenology in free-fermion measurement-only models~\cite{Skinner2019}.
In addition, the dual area-law phases in these models may be characterized by different types of quantum order.

A distinctive aspect of dynamics involving projective measurements is the possibility of infinite entanglement velocity, or `spooky action at a distance'. In Sec.~\ref{sec:loc} we have introduced a new probe of information spreading, built from the mutual information between parts of the system and an entangled reference qubit; surprisingly, both in the volume-law phase and at criticality we have found that entanglement spreads within a ballistic lightcone, suggesting the statistical emergence of locality. 

The volume-law entangled phase in these models is a potentially distinctive type of random quantum error-correcting code, as we discussed in Sec.~\ref{sec:code}. 
An ensemble of measurements, implemented randomly, protects quantum information from \emph{any} future sequence of measurements drawn randomly from that same ensemble, without any other outside intervention.
This could be used to design tailored quantum codes:
often times in quantum computing devices, noise is not uniformly random, but has biases and correlations; by selecting the measurement ensemble to reflect the detailed properties of the noise, it may be possible to  tailor codes for the specific noise configuration of the device~\cite{Tuckett19}.

Concrete connections between fault-tolerant quantum computation and this family of measurement-only dynamics arise in the context of topological quantum error correcting codes \cite{Kitaev03}.  
Topological codes are one of the leading candidates for realizing scalable quantum computing\cite{Fowler12}.  At a practical level, implementing a quantum memory with such a code amounts to repeated rounds of multi-site, local Pauli measurements to detect errors\cite{Dennis02,Kelly15}.  Recovery operations are typically implemented after multiple rounds of measurements to avoid errors in syndrome extraction, which results in a $d+1$-dimensional  quantum nonequilibrium problem similar to the type studied here. 
In the ideal scenario, these measurements are all commuting with each other, but a natural error model is to allow for these measurements to become noncommuting with some probability due to unitary gate errors that occur during multi-site measurements.   The threshold analysis of this model maps exactly to a MOM.  
As a result, some of the insights obtained from studying measurement-only dynamics in stochastic, unstructured settings may prove useful in the threshold and decoding analysis of such topological codes.  
More ambitiously, it may be that the dynamics introduced here can be naturally realized in a fault-tolerant manner in NISQ devices through small changes to experimental setups designed  to implement  topological quantum error correction.

Our work points to several interesting directions for future research.
The entanglement critical points we have discovered raise many questions. The critical exponents for the volume- to area-law transition appear different from those that were previously found in generic unitary-projective circuits, suggesting that these transitions might belong to a different universality class; more intensive numerical work needs to be done to settle this question. 
The nature of the area-law to area-law critical points in bipartite ensembles also remains unclear, and particularly whether they can be mapped to percolation in a suitable loop model, like in the free-fermion case~\cite{Nahum2019}.
Strikingly, the critical points for these MOM phase transitions also appear to be CFTs, despite no apparent space-time symmetry. 
It would be interesting to understand if this is a consequence of a mapping to classical statistical mechanical models, similar to the case of unitary-projective dynamics~\cite{Jian2019}.  
While the mapping to a statistical mechanics model remains unclear for hybrid circuits in the Clifford case~\cite{LiFisher2020}, 
adapting the construction of Ref.~[\onlinecite{Jian2019}] to study these transitions in the Haar-random limit would be a natural extension.


\acknowledgments
We thank Roderich Moessner, Jed Pixley, Romain Vasseur, Dominic Williamson, Justin Wilson and Aidan Zabalo for insightful discussions and collaborations on related topics.
M.I. and D.A.H. were supported with funding from the Defense Advanced Research Projects Agency (DARPA) via the DRINQS program. The views, opinions and/or findings expressed are those of the authors and should not be interpreted as representing the official views or policies of the Department of Defense or the U.S. Government. 
M.I. was also funded in part by the Gordon and Betty Moore Foundation's EPiQS Initiative through Grant GBMF4302 and GBMF8686.
S.G. acknowledges support from NSF DMR-1653271. 
V.K. acknowledges support from the US Department of Energy, Office of Science, Basic Energy Sciences, under Early Career Award No. DE-SC0021111. 
D.A.H. was also supported in part by a Simons Fellowship.


\appendix

\section{Hybrid $l$-bit circuit as a measurement-only model \label{app:mom_lbit}}

Here we define a MOM based on the unitary-projective $l$-bit circuit introduced in Sec.~\ref{sec:lbit} and discuss its phenomenology in detail.


\subsection{Ensemble}
The $l$-bit unitary-projective circuits discussed in Sec.~\ref{sec:lbit}, with $p_z=0$ (i..e measurements only along $X$) naturally lead to measurement-only dynamics specified by the ensemble
\begin{equation}
\mathcal E_{l\text{-bit}} = \left\{ O_{\boldsymbol \alpha} = X_0 \prod_{\ell = 1-n}^{n-1} Z_\ell^{\alpha_\ell} \right\}
\label{eq:lbit_momensemble}
\end{equation}
where $\boldsymbol\alpha \in \{0,1\}^{2n-1}$ labels the operator species (we omit phase factors for simplicity, writing $XZ$ {\it in lieu} of $Y$).
All operators are characterized by a `central site' that is either a Pauli $X$ or $Y$ (which corresponds to the site where an $X$ measurement is made in the hybrid circuit, possibly followed by a phase gate) and tails on both sides that are made exclusively of $\mathbb I$ or $Z$ Pauli matrices, with equal probability of $1/2$ (which arise from the $\CZ$ gates). 

As written, the above ensembles have range $r=2n-1$; however, any two operators displaced by $|\ell| \geq n$ commute, since in that case only their tails (made entirely of $\mathbb I$ and $Z$) overlap. 
Thus the effective range, as specified by the connectivity of the frustration graph, is $\tilde{r} = n$. 


\subsection{Connection with l-bit circuit model}

The original unitary-projective $l$-bit circuit has a measurement rate, $p$, that (once translated to the measurement-only language) tends to make the measurements more commuting, and thus drives the dynamics towards the area-law phase. 
In the original circuit, if two measurements take place in the same time slice within distance $n$ of each other, they manifestly commute;
this commutation must be maintained even after getting rid of the $\CZ$ gates via the ``trick'' in Eq.~\eqref{eq:trick}.
This means that, when switching from the unitary-projective to the measurement-only pictures, an amount of correlation (or memory) is built into the measurements drawn from the ensemble: for instance, after drawing $X_{i} Z_{i+1}$ one is more likely to draw the commuting observable $Z_i X_{i+1}$ than the anticommuting observable $X_{i+1}$.
In dropping such correlations, we are implicitly taking a $p\to 0^+$ limit (and concurrently rescaling time, since the unit of time in measurement-only dynamics has on average $O(1)$ measurements per site, as opposed to $O(p)$ measurements per site per layer in the unitary-projective circuit). 
This makes the relation between two models (unitary-projective and measurement-only) a bit subtle.
The $p\to 0^+$ limit (which more accurately stands for $pL\ll 1$, i.e. very low probability of having two measurements next to each other in the same layer of the original circuit) is meaningless when taking the thermodynamic limit $L\to\infty$ first.
However, reintroducing the correlations mentioned above can only push the dynamics towards area law, so the measurement-only dynamics provides a strict upper bound to the steady-state entanglement of the hybrid circuit with any finite $p$. 


\subsection{Entanglement transition}

As seen in Sec.~\ref{sec:lbit}, the unitary-projective $l$-bit dynamics 
admit a volume-law phase for integer $n\geq 4$, while $n\leq 3$ is area-law independent of measurement rate $p$. 
To locate the phase boundary, one must continue $n$ to fractional values, $n=n^\star+\epsilon$ ($n^\star \in \mathbb{N}$, $0\leq \epsilon<1$). 
In the hybrid $l$-bit circuit of Sec.~\ref{sec:lbit} we do so by acting with $\CZ$ gates between qubits $i,j$ with probability 
\begin{equation}
\text{Prob}(\CZ_{ij}) = \left\{ 
\begin{aligned}
1/2 & \text{ if } |i-j|<n^\star \\
\epsilon/2 & \text{ if } |i-j|=n^\star \\
0 & \text{ if } |i-j|>n^\star 
\end{aligned} 
\right.
\label{eq:app_probCZ}
\end{equation}
Clearly $\epsilon = 0$ and $\epsilon=1$ yield the model with integer $n=n^\star$ and $n=n^\star+1$ respectively.
The fractional values of $n$ used in Fig.~\ref{fig:lbitphasediag}(b) are defined in this way.

Taking the measurement rate $p\to 0^+$ in the above-defined hybrid models always returns a MOM with integer $n$, specifically the model in Eq.~\eqref{eq:lbit_momensemble} with $n = n^\star+1$: 
measurements become so infrequent that \emph{any allowed gate} $\CZ_{ij}$ acts a large number of times in between consecutive measurements, and thus the corresponding exponent $\alpha_\ell$ ($\ell=i-j$) in Eq.~\eqref{eq:lbit_momensemble} is equally likely to be 0 or 1. 
This is the reason why the phase boundary in the hybrid circuit's entanglement phase diagram (Fig.~\ref{fig:lbitphasediag}) approaches $n=3^+$ as $p\to 0^+$: for all $0<\epsilon<1$, the hybrid circuits with $n=3+\epsilon$ map onto the same MOM with $n=4$. 

The family of $l$-bit-inspired MOMs, Eq.~\eqref{eq:lbit_momensemble}, can be independently continued to fractional $n = n^\star + \epsilon$ by defining
\begin{equation}
O_{\boldsymbol \alpha} = X_0 \prod_{\ell = -n^\star}^{n^\star} Z_\ell^{\alpha_\ell}\;, \qquad
P_{\boldsymbol\alpha} = \frac{1}{\mathcal N} \epsilon^{\alpha_{-n^\star}+\alpha_{n^\star}} \;,
 \label{eq:lbitMOMensemble}
\end{equation}
where $\mathcal N$ is a probability normalization.
In other words, $Z$ operators at the extreme points of the tail, $\ell = \pm n^\star$, are less frequent than those at other points by a factor of $\epsilon$.

When sweeping $n$ from 3 to 4 in the above models, we encounter an entanglement transition surprisingly close to $n=3$:
the critical point is estimated at $n_c = 3.020(3)$.
We emphasize that this critical point is incompatible with $n=3$, where the dynamics unambiguously converges to an area-law. 
Nonetheless, proximity to this critical point endows the $n=3$ $l$-bit model with a very long, though finite, correlation length.

\begin{figure}
\centering
\includegraphics[width=\columnwidth]{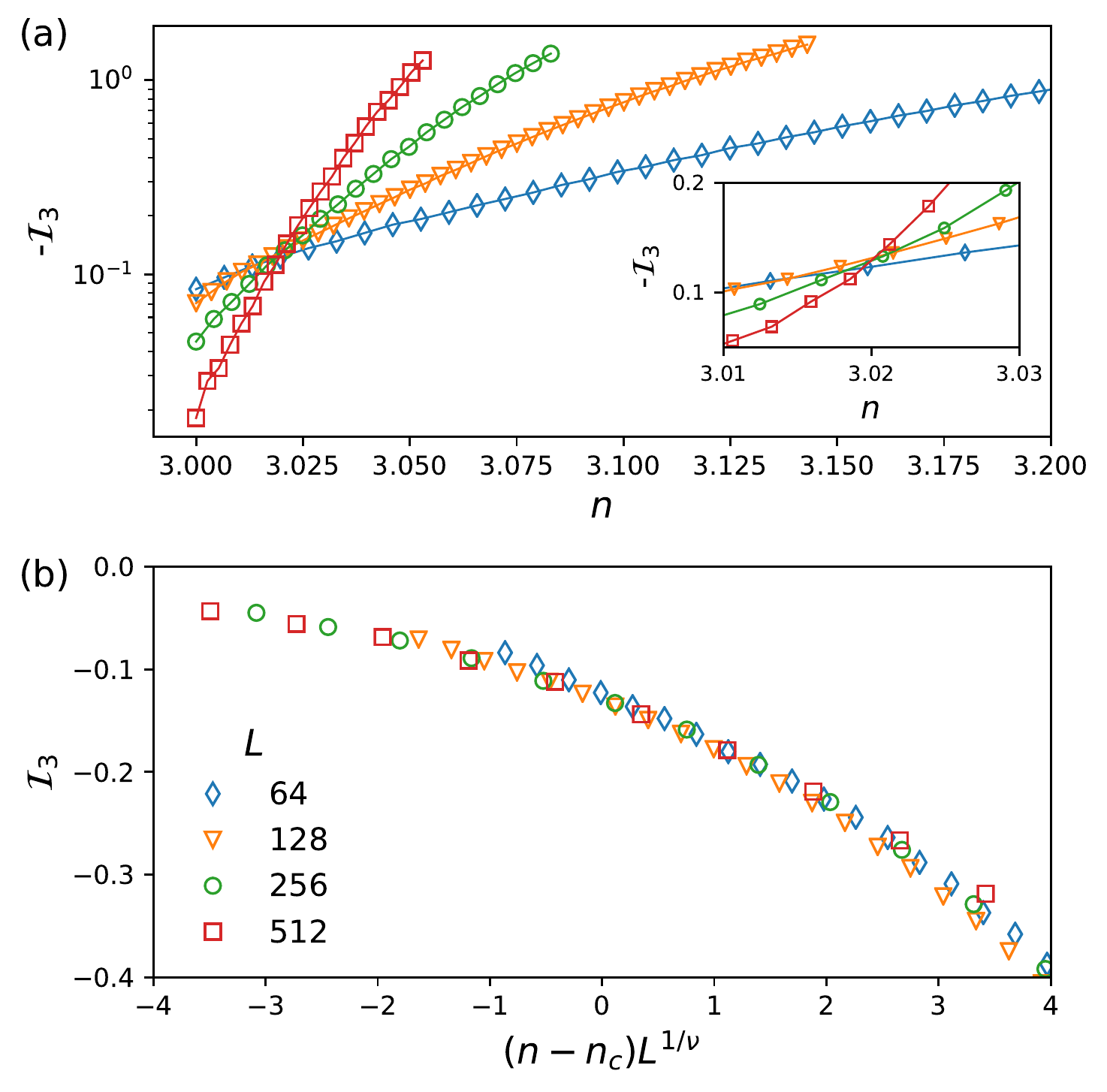}
\caption{Entanglement transition in the $l$-bit MOM continued to ``fractional $n$'' between 3 and 4.
(a) Tripartite mutual information $\tmi$ as a function of $n$.
Different sizes $64\leq L \leq 512$ show a crossing at $n = n_c = 3.020(3)$ (inset). 
(b) Scaling collapse of the data with exponent $\nu = 1.1$. 
\label{fig:lbit_transition}}
\end{figure}

We repeat the analysis of Sec.~\ref{sec:critical} for this model. 
Here, too, we find a continuous transition with logarithmic entanglement entropy, $S (\ell) = K \ln \ell$ ($K = 0.8(1)$); 
dynamical exponent $z=1$; 
and (see Fig.~\ref{fig:lbit_transition}) correlation length critical exponent $\nu = 1.1(1)$. The critical properties are thus consistent with those of the model examined in Sec.~\ref{sec:critical}.


\section{Details on stabilizer dynamics \label{app:stab_rules}}

Here we review the stabilizer formalism for Pauli measurements to complement the discussion in Sec.~\ref{sec:stab_rules}.
Let us consider a pure stabilizer state, as in Eq.~\eqref{eq:rho_stabilizer} with $S=0$.
Under Clifford unitaries and Pauli measurements, the list of stabilizer generators can be updated in polynomial time with operations that amount to linear algebra over $\mathbb Z_2$~\cite{Aaronson2004}; here we review the update rules.

Because all stabilizer generators must commute with one another, $[g_i,g_j]=0$ (as it is impossible for two anticommuting operators to share a $+1$ eigenstate), three possibilities arise when a Pauli string $O$ is measured:
\begin{enumerate}[label={\bf \arabic*}]
\item $O$ anticommutes with exactly one generator, say $g_1$. The measurement outcome is $\sigma =\pm 1$ chosen randomly; $g_1$ is updated to $g_1' = \sigma O$. 
\item $O$ anticommutes with several generators, say $\{g_1, \dots g_k\}$. 
This case can always be reduced to the previous one by a gauge transformation, i.e. redefinition of the generators (for example $g_i' = g_i g_1$ for all $1<i \leq k$, after which only $g_1$ anticommutes with $O$).
\item $O$ commutes with all generators $\{g_i\}$. In a pure state, this guarantees that $O$ is a stabilizer, i.e. it can be written as a product of $g_i$'s (up to a sign). The measurement outcome is deterministic and the state is unchanged by the measurement. 
\end{enumerate} 

It can also be useful to adopt the `purification' point of view~\cite{Gullans2019A}, where one starts with a mixed stabilizer state (Eq.~\eqref{eq:rho_stabilizer} with $S>0$) represented by an incomplete list of generators $\{g_i:\ i=1,\dots L-S\}$, possibly an empty list $\{\}$ for the maximally mixed state $\rho =\mathbb I/2^L$ ($S=L$).
Based on the time it takes the state to purify one can define purification phases:
a `mixed phase' where the state remains mixed for exponentially long times, and a `pure phase' where the state becomes pure in time $O(\log L)$.
These purification phases correspond to the entanglement phases for pure states (mixed $\leftrightarrow$ volume-law, pure $\leftrightarrow$ area-law).
In this mixed-state scenario, cases 1 and 2 from the list above play out in the same way, but case 3 must be subdivided into
\begin{enumerate}[label={\bf 3\Alph*}]
\item $O$ commutes with all generators but is \emph{not} part of the stabilizer group (i.e., it is a logical operator). The measurement outcome is $\sigma = \pm 1$, chosen at random, and a new generator $g_{L-S+1} = \sigma O$ is added to the list. The state loses one bit of entropy ($S\mapsto S-1$).
\item $O$ commutes with all generators and is part of the stabilizer group (up to a sign). The measurement outcome is deterministic, and the state is unchanged by the measurement. 
\end{enumerate}


\section{Details on frustration \label{app:frustration}}

In this Appendix we discuss the frustration graph and tensor, introduced in Sec.~\ref{sec:frustration}, in greater detail.
In Sec.~\ref{app:f_simulation} we show that the frustration graph contains all the information needed to simulate measurement-only purification dynamics (provided operators in the ensemble are algebraically independent).
Using this fact, in Sec.~\ref{app:f_equivalence} we discuss how the frustration graph can show that seemingly different models have equivalent purification dynamics, and thus equivalent entanglement/purification phase diagrams.
Finally in Sec.~\ref{app:f_factor} we discuss the frustration of factorizable ensembles introduced in Sec.~\ref{sec:factorizable}.


\subsection{Simulation of the dynamics\label{app:f_simulation}}

The stabilizer update rules reviewed in Appendix~\ref{app:stab_rules} do not require specific knowledge of the operators $O_\alpha$.
Rather, they only depend on their mutual anticommutation and algebraic dependence properties.  
Given these data, one can simulate the purification dynamics and determine the purification phase diagram.

Consider measuring an operator $O_{\alpha,n}$ (species $\alpha$, position $n$).
To update the stabilizer state, we must first test the commutation between this operator and the existing generators $\{ g_i \}$.
This only requires knowledge of the frustration tensor $\Gamma^{\alpha\beta}_\ell$.
Indeed, let us decompose each generator as (up to a phase) 
\begin{equation}
g_i = \prod_{\alpha, n} O_{\alpha,n}^{v^i_{\alpha, n}}
\label{eq:stab_parametrization}
\end{equation}
for appropriate coefficients $v_{\alpha,n}^i \in \mathbb Z_2$, not necessarily unique (such a decomposition is known, trivially, for the initial maximally mixed state $\rho \propto \mathbb I$, and can be consistently updated without full knowledge of the $O_\alpha$, as we show next).
It is convenient to introduce the ``scalar commutator'' between Pauli strings $A\circ B\in \mathbb Z_2$, $A\circ B = 0$ if $A,B$ commute, 1 otherwise.
The operators $O_{\alpha,n}$ and $g_i$ commute if and only if
\begin{equation}
O_{\alpha, n} \circ g_i = \sum_{\beta, n'}  \Gamma^{\alpha,\beta}_{n-n'} v_{\beta, n'}^i \equiv 0 \mod 2 \;.
\end{equation}
These bits determine which of the cases in Appendix~\ref{app:stab_rules} is realized.
\begin{enumerate}[label={\bf \arabic*}]
\item If only $g_1$ anticommutes with $O_{\alpha, n}$, then we update $v^1_{\alpha',n'} \mapsto \delta_{\alpha,\alpha'} \delta_{n,n'}$ (i.e. discard $g_1$ and replace it with $O_{\alpha,n}$).
\item  Similar, but one must first redefine all the anticommuting generators $g_2, \dots g_k$ according to $g_i\mapsto g_i' = g_1 g_i$, which is done as $v^i_{\alpha,n} \mapsto v^i_{\alpha,n} + v^1_{\alpha,n}$ (modulo 2).
\end{enumerate}
Case {\bf 3} (no anticommutation) requires distinguishing between {\bf 3A} (measurement of a logical operator) and {\bf 3B} (measurement of a stabilizer).
This cannot be done in general without information about algebraic dependence between the operators.
If we assume that all operators in the ensemble are independent (often the case with two operator species), testing $O_{\alpha,n}$'s algebraic dependence on the $\{g_i\}$ is equivalent to testing linear independence of the new $\mathbb{Z}_2$ vector $v^\text{new}_{\alpha',n'} \equiv \delta_{\alpha,\alpha'}\delta_{n,n'}$ on all existing $\mathbb{Z}_2$ vectors $\{v^i\}$, which reduces to a linear algebra problem over $\mathbb{Z}_2$. 
\begin{enumerate}[label={\bf 3\Alph*}]
\item If $v^\text{new}$ is linearly independent from the set $\{v^i: i=1,\dots L-S\}$, it gets added as a new stabilizer, $v^{L-S+1} \equiv v^\text{new}$
\item Otherwise, nothing happens
\end{enumerate}

In conclusion, all the data needed to update the stabilizer generators for a given measurement is contained in the frustration tensor $\Gamma^{\alpha\beta}_\ell$ and any algebraic dependence relations between ensemble operators.
This is sufficient to simulate purification dynamics and thus decide the purification phase.
None of this requires explicit knowledge of the $\{O_{\alpha,n} \}$ operators.


\subsection{Equivalence between ensembles\label{app:f_equivalence}}

If two ensembles have frustration graphs that can be transformed into one another by moving vertices around without breaking or creating any edges, then there is a mapping $O_{\alpha,n} \leftrightarrow O_{\beta,m}'$ between the operators in the two ensembles which preserves all commutation relations. 
If the probability distributions are invariant under this mapping as well (always the case for the uniform distribution), then the purification dynamics induced by the two ensembles are equivalent.

Among the applications of these graph-theoretic ideas is a method to determine whether an ensemble of Pauli strings is equivalent to free fermion measurements~\cite{Chapman2020}.
For free fermion ensembles, the frustration graph is a `line graph': there exists another graph whose vertices are Majorana fermions and whose edges are the ensemble operators. 
The property of being a line graph can be tested in time $O(L)$ in the general case, and in time $O(r)$ for our local (range $r$), translationally-invariant models.

\begin{figure}
\centering
\includegraphics[width=0.99\columnwidth]{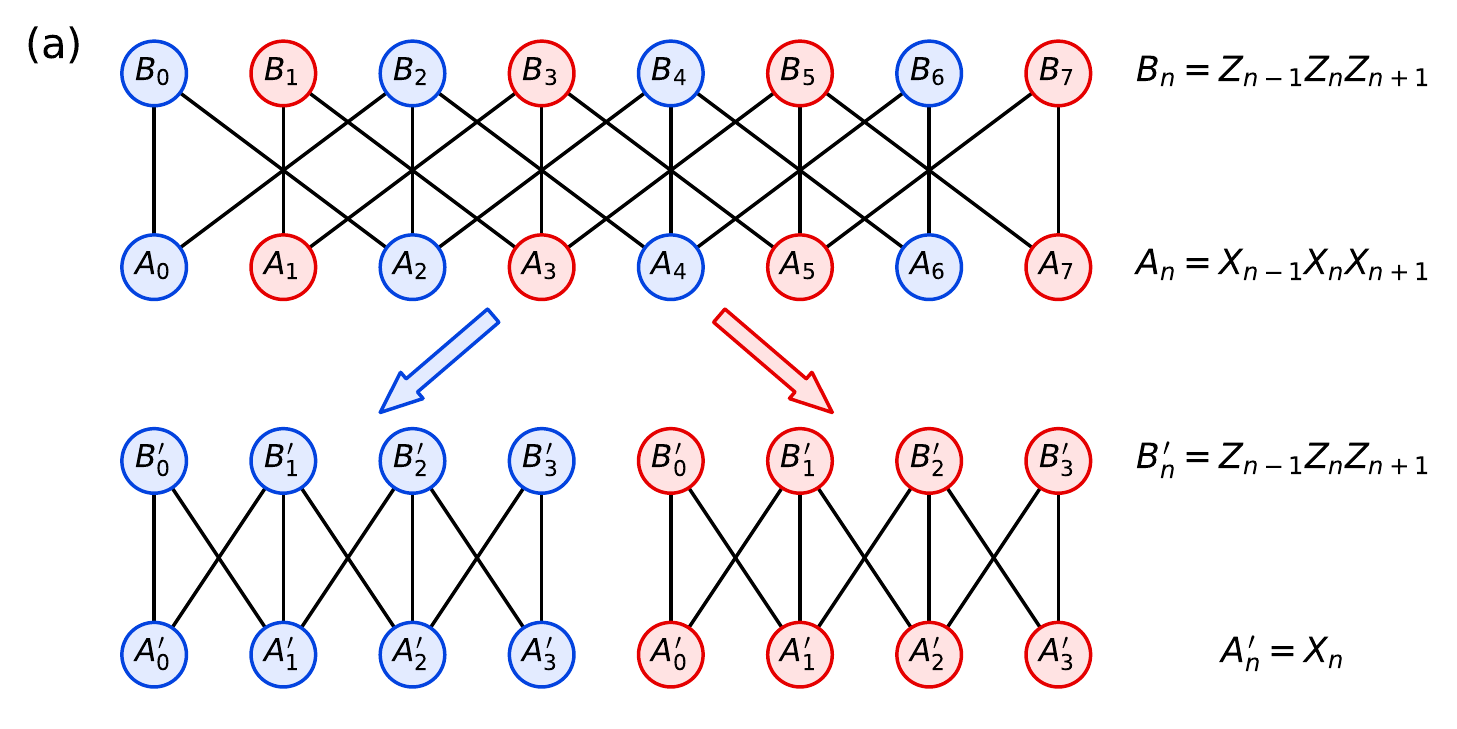}\\
\includegraphics[width=0.99\columnwidth]{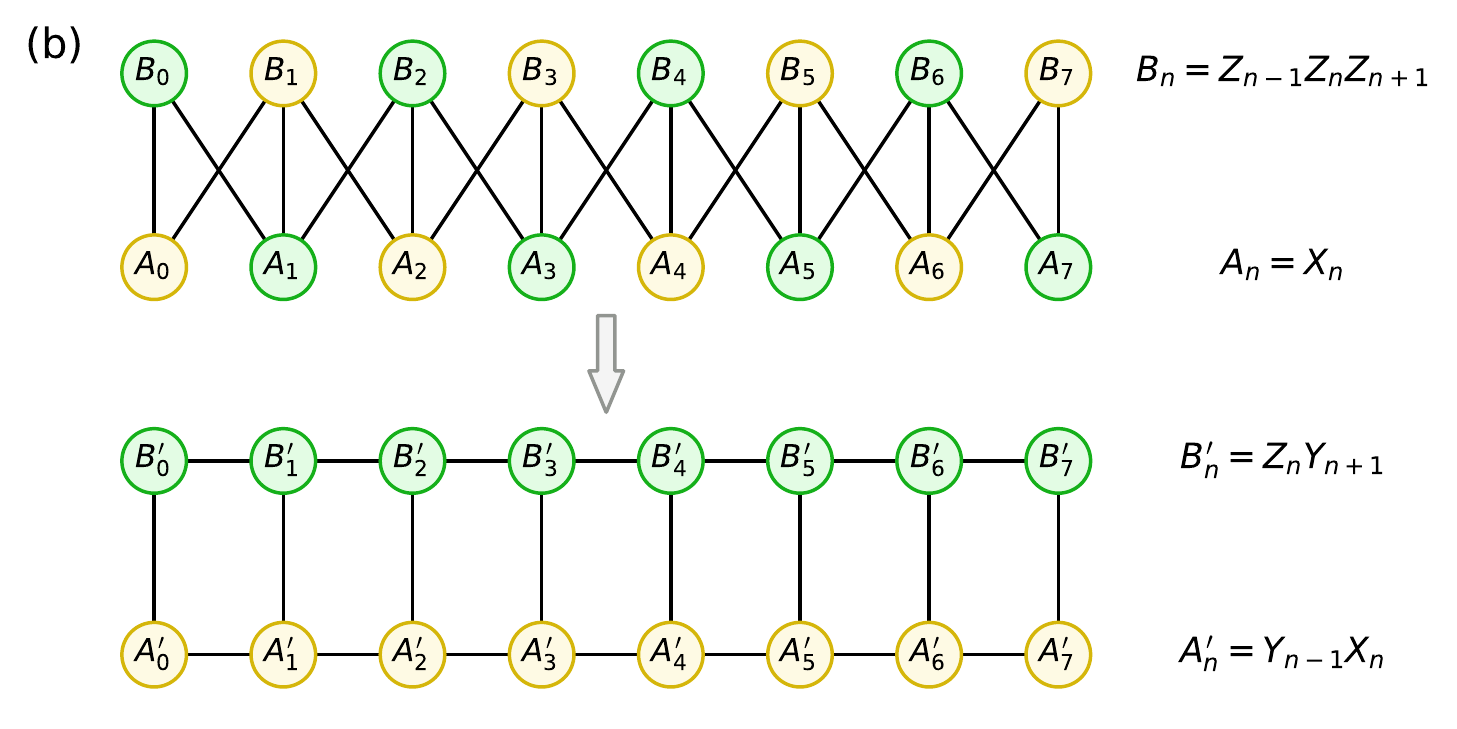}
\caption{Frustration graph reveals equivalence between different measurement ensembles. 
(a) The frustration graph for the ensemble ${\mathcal E}_1 = \{ Z_0Z_1Z_2, X_0X_1X_2\}$ has two connected components, highlighted by the color scheme. Separating the two yields two copies of ensemble ${\mathcal E}_2 = \{Z_0Z_1Z_2,X_1\}$.
(b) The ensemble ${\mathcal E}_2$ is in turn equivalent to $\mathcal E_3 = \{Z_0 Y_1, Y_0X_1\}$ via a permutation of vertices, as highlighted by the color scheme. 
\label{fig:frustration}}
\end{figure}

Aside from mappings to free fermions, graph equivalence allows us to prove that seemingly distinct ensembles belong to the same phase, or even more detailed relations between their steady-state entanglement entropy.
As an example, we show in Fig.~\ref{fig:frustration} that the ensembles $\mathcal E_1 = \{X_0 X_1 X_2, Z_0 Z_1 Z_2 \}$, $\mathcal E_2 = \{X_1, Z_0Z_1Z_2\}$ and $\mathcal E_3 = \{ X_0 Z_1, Z_0 Y_1\}$ (with the two species sampled uniformly in all cases) are all in the same phase:
the graph for $\mathcal E_1$ splits into two subgraphs, each of which is equivalent to $\mathcal E_2$; additionally, $\mathcal E_2$ and $\mathcal E_3$ are equivalent to each other.

\subsection{Frustration of factorizable ensembles \label{app:f_factor}}

Here we discuss the frustration of factorizable ensembles, Sec.~\ref{sec:factorizable} and how it relates to the on-site probability distribution of Pauli matrices, $\mathbf{q} = (q_X, q_Y, q_Z)$.

These models have a large number of operators species, $3^r$, which makes the frustration graph itself large and not useful.
More useful information can be gleaned from the \emph{averaged} frustration tensor,
\begin{equation}
\overline{\Gamma}_\ell \equiv \sum_{\alpha,\beta} P_\alpha P_\beta \Gamma_\ell^{\alpha \beta}\;,
\end{equation}
which captures the probability that two operators drawn at random from the ensemble anticommute, as a function of their spatial displacement $\ell$.

Let us start by considering a displacement of $\ell = r-1$, i.e. Pauli strings overlapping on a single site.
The probability that they anticommute is by 
$$
\overline{\Gamma}_{r-1} = 2(q_X q_Y + q_Y q_Z + q_Z q_X) = 2 \mathbf q \cdot R_{2\pi/3}^{(111)} \mathbf q \;,
$$
where $R_\theta^{(111)}$ is a rotation about the $(111)$ axis in $\mathbf q$ space. 
Decomposing $\mathbf q = \mathbf q_0 + \delta \mathbf q$, with $\mathbf q_0 = (1,1,1)/3$ and $\delta \mathbf q \cdot \mathbf q_0 = 0$ (fixed by the normalization of probabilities), we obtain 
$$
\overline{\Gamma}_{r-1}  = 2q_0^2 + 2 \delta q^2 \cos \frac{2\pi}{3} = \frac{2}{3} - \delta q^2 \;.
$$
Values of $\bar{\Gamma}_\ell$ for $\ell < {r}-1$ are found recursively:
$\bar{\Gamma}_{r-k}$ is the probability that two strings overlapping on $k$ sites anticommute; splitting the overlapping region into two intervals of length $k-1$ and 1 gives
\begin{align*}
\overline{\Gamma}_{r-k} 
& = \overline{\Gamma}_{r-1}(1-\overline{\Gamma}_{r-k+1}) + \overline{\Gamma}_{r-k+1}(1-\overline{\Gamma}_{r-1}) \\
& = \overline{\Gamma}_{r-k+1} (\delta q^2 - 1/3) + (2/3-\delta q^2) \;,
\end{align*}
which can be turned into a geometric series for $1/2-\overline{\Gamma}_{r-k}$, yielding
\begin{equation}
\overline{\Gamma}_\ell = \frac{1}{2} - \frac{1}{2} \left(2\delta q^2- \frac{1}{3} \right)^{r-\ell}
\end{equation}
This shows that the average anticommutation probabilities for these ensembles are fixed by the radial distance $\delta q$ from the center of parameter space.
This offers a qualitative explanation for the approximately-circular phase boundaries in Fig.~\ref{fig:triangles}.
At $\delta q^2 = 2/3$ (edges of the triangle), we have $\overline{\Gamma}_\ell\equiv 0$, i.e. a fully un-frustrated ensemble, as expected; 
decreasing $\delta q$ increases anticommutation.

\bibliography{MOM}

\end{document}